\pdfoutput=1
\documentclass[fleqn,useAMS,usenatbib]{mnras}
\usepackage[T1]{fontenc}
\usepackage{ae,aecompl}
\usepackage{mathptmx} 
\usepackage{amsmath} 
\usepackage{amssymb} 

\usepackage{graphicx}	
\usepackage{color}		
\usepackage[usenames,dvipsnames]{xcolor} 
\usepackage{ifthen}		

\newcommand{\masyr}{mas\,yr$^{-1}$}
\newcommand{\kms}{\textrm{km\,s$^{-1}$}}
\newcommand{\vsini}{\ensuremath{v\sin i}}
\newcommand{\msun}{$M_{\sun}$}

\newcommand{\starlong}[1]{%
    \ifthenelse{\equal{#1}{71}}{2MASS J10284580--2830374}{}%
    \ifthenelse{\equal{#1}{74}}{2MASS J10055826--2131142}{}%
    \ifthenelse{\equal{#1}{77}}{2MASS J10491880--2509235}{}%
    \ifthenelse{\equal{#1}{78}}{2MASS J10585054--2346206}{}%
    \ifthenelse{\equal{#1}{82}}{2MASS J12002750--3405371}{}%
    \ifthenelse{\equal{#1}{83}}{2MASS J12023799--3328402}{}%
    \ifthenelse{\equal{#1}{91}}{2MASS J14224891--3623009}{}%
    \ifthenelse{\equal{#1}{205}}{2MASS J12392312--5702400}{}%
}
\newcommand{\starshort}[1]{%
    \ifthenelse{\equal{#1}{71}}{2M1028--2830}{}%
    \ifthenelse{\equal{#1}{74}}{2M1005--2131}{}%
    \ifthenelse{\equal{#1}{77}}{2M1049--2509}{}%
    \ifthenelse{\equal{#1}{78}}{2M1058--2346}{}%
    \ifthenelse{\equal{#1}{82}}{2M1200--3405}{}%
    \ifthenelse{\equal{#1}{83}}{2M1202--3328}{}%
    \ifthenelse{\equal{#1}{91}}{2M1422--3623}{}%
    \ifthenelse{\equal{#1}{205}}{2M1239--5702}{}%
}

\title[New members of TW Hydrae and Sco-Cen]{New members of the TW~Hydrae Association and two accreting M-dwarfs in Scorpius-Centaurus}
\author[Murphy, Lawson \& Bento]{Simon~J.~Murphy$^{1,2}$\thanks{E-mail: simon.murphy@anu.edu.au}, Warrick~A.~Lawson$^2$ and Joao~Bento$^1$\\
$^{1}$ Research School of Astronomy \& Astrophysics, Australian National University, Canberra, ACT 2611, Australia\\
$^{2}$  School of Physical, Environmental and Mathematical Sciences, University of New South Wales Canberra, ACT 2600, Australia}

\pubyear{2015}

\begin{document}
\label{firstpage}
\pagerange{\pageref{firstpage}--\pageref{lastpage}}
\maketitle

\begin{abstract}
We report the serendipitous discovery of several young mid-M stars found during a search for new members of the 30--40~Myr-old Octans Association. Only one of the stars may be considered a possible Octans(-Near) member. However, two stars have proper motions, kinematic distances, radial velocities, photometry and \ion{Li}{i} $\lambda$6708 measurements consistent with membership in the 8--10~Myr-old TW Hydrae Association. Another may be an outlying member of TW Hydrae but has a velocity similar to that predicted by membership in Octans. We~also identify two new lithium-rich members of the neighbouring Scorpius-Centaurus OB Association (Sco-Cen). Both exhibit large 12 and 22~\micron\ excesses and strong, variable H$\alpha$ emission which we attribute to accretion from circumstellar discs. Such stars are thought to be incredibly rare at the $\sim$16~Myr median age of Sco-Cen and they join only one other confirmed M-type and three higher-mass accretors outside of Upper Scorpius. The serendipitous discovery of two accreting stars hosting large quantities of circumstellar material may be indicative of a sizeable age spread in Sco-Cen, or further evidence that disc dispersal and planet formation time-scales are longer around lower-mass stars. To aid future studies of Sco-Cen we also provide a newly-compiled catalogue of 305 early-type \emph{Hipparcos} members with spectroscopic radial velocities sourced from the literature.
\end{abstract}

\section{Introduction}\label{sec:introduction}

Nearby stars with ages of 5--20~Myr are ideal laboratories in which to investigate the later stages of star and planet formation. It is on these time-scales that the gas-rich discs which feed giant planet formation and circumstellar accretion dissipate and are replaced by dusty debris discs replenished by the collisions of rocky bodies \citep{Wyatt08,Williams11}. Such stars are also prime targets for direct imaging observations of young, bright exoplanets and substellar companions \citep[e.g.][]{Marois08,Lagrange10}. The discovery of young moving groups in the  solar neighbourhood over the last two decades \citep[see reviews by][]{Zuckerman04a,Torres08} has enabled the study of circumstellar material and processes in this important age range at high spatial resolution and sensitivity across a wide range of stellar masses. 

Foremost amongst the nearby moving groups is the retinue of coeval pre-main sequence stars \citep[age 8--10~Myr;][]{Weinberger13,Ducourant14} within a few tens of degrees and comoving with the T Tauri star TW Hydra \citep{Rucinski83}. The so-called TW Hydrae Association \citep[TWA;][and references therein]{de-la-Reza89,Kastner97,Torres08,Schneider12b} now includes approximately 30 systems, spanning spectral types of A0 through to brown dwarfs. Its members were identified by a variety of means, most commonly as common proper motion sources detected in X-rays by the \emph{ROSAT} mission \citep[e.g.][]{Webb99,Zuckerman01b,Song03}, but also through near-IR photometry \citep{Gizis02b,Looper07,Looper10}, circumstellar disc excesses \citep{Gregorio-Hetem92,Schneider12} and activity \citep{Shkolnik11}.  The current membership of TWA is dominated by K and M-type stars, although there appears to be a dearth of members later than $\sim$M3 \citep{Shkolnik11}. This is  partly a consequence of the heavy reliance of previous searches on the \emph{ROSAT} All Sky Survey \citep{Voges99}, which, for the saturated coronal X-ray emission commonly seen in young stars ($L_{X}/L_{\text{bol}}\approx-3$), had a limiting spectral type of around M2 at the $\sim$60~pc mean distance of TWA. Recent efforts  exploiting deeper UV photometry from the \emph{GALEX} mission \citep{Martin05} have proven successful at identifying young stars from their enhanced chromospheric activity \citep{Shkolnik11,Rodriguez11,Murphy15}, especially in conjunction with modern photometric and proper motion surveys. 

\begin{figure*} 
   \centering
   \includegraphics[width=0.995\linewidth]{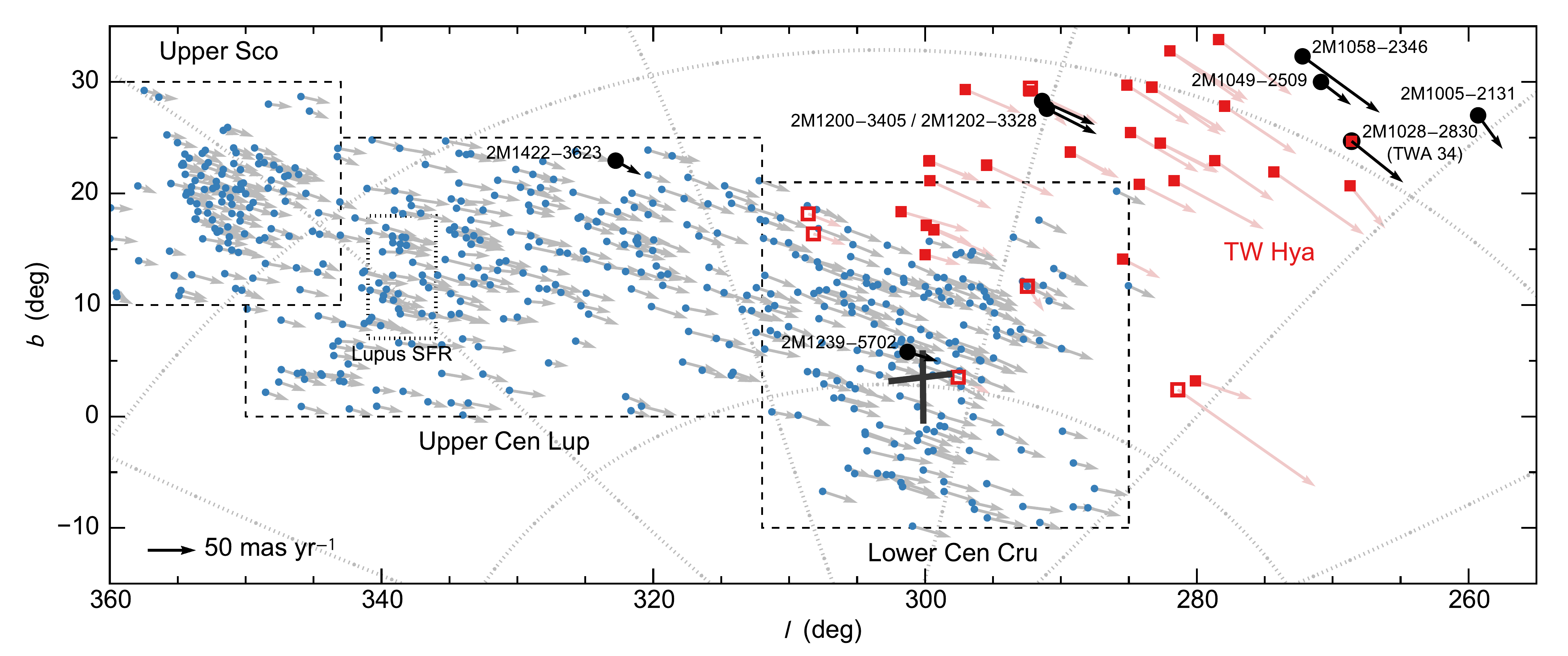} 
   \caption{Galactic positions and proper motions of the eight lithium-rich stars identified in this work (black points). Also plotted are proposed Sco-Cen   \citep[blue dots;][]{de-Zeeuw99} and TWA members \citep[red squares;][]{Ducourant14}. Open TWA symbols denote non-members, most of which are probably background LCC stars. The star near $l=280$~deg with the large proper motion is TWA~22, a likely $\beta$ Pictoris member \citep{Teixeira09}. For reference, the well-known Southern Cross asterism is also plotted at $l=300$~deg.}  \label{fig:skyplot}
\end{figure*}

Immediately south of TWA lies the Scorpius-Centaurus OB Association \citep[Sco-Cen, Sco OB2;][]{Blaauw64,de-Zeeuw99,Preibisch08}, the closest site of recent massive star formation and the dominant population of pre-main sequence stars in the solar neighbourhood. The association has been implicated in the formation of several nearby young moving groups, including TWA \citep{Mamajek01,Fernandez08,Murphy13}. Sco-Cen has traditionally been divided into three subgroups (Fig.~\ref{fig:skyplot}), each with subtly different mean distances, ages and velocities: Upper Scorpius (US, 145~pc), Upper Centaurus Lupus (UCL, 140~pc) and Lower Centaurus Crux \citep[LCC 120~pc;][]{de-Zeeuw99}. The subgroups have median ages of approximately 10~Myr, 16~Myr and 17~Myr, respectively \citep{Mamajek02,Pecaut12}. While the full mass function of US is well known   \citep[e.g.][]{Preibisch98,Lodieu11,Rizzuto15}, the older and more southern UCL and LCC subgroups are under-studied below 1~\msun. Only $\sim$100 K~and M-type members are known, mostly detected by \emph{ROSAT} \citep{Mamajek02,Preibisch08,Song12} but also more recently by \emph{GALEX} \citep{Rodriguez11}. Any reasonable mass function predicts several thousand low-mass members awaiting discovery across Sco-Cen.   

In this contribution we describe the serendipitous discovery and spectroscopic confirmation of one existing and up to three new M-type members of TWA, as well as two previously unknown members of Sco-Cen exhibiting signs of accretion from circumstellar discs. These six stars, and another two spectroscopically-young stars not yet attributable to any known group, were identified during an ongoing search for members  of the 30--40~Myr-old Octans Association \citep{Torres08,Murphy15}.  In the following section we briefly describe the candidate selection procedures and observations, deferring a full description of the Octans survey to an upcoming work (Murphy et al. 2015, in preparation).

\section{Candidate selection and observations}

The aim of our photometric and proper motion survey was to identify new  Octans members on the cool side of its lithium depletion boundary (LDB) in order to better refine the age of the association. Candidates for spectroscopy were selected from the SPM4 proper motion catalogue \citep{Girard11} in a similar manner to \citet{Murphy15}, with initial cuts of $V_{\textrm{SPM4}}<20$, $V-K>5$, $J-H<0.7$ and proper motion errors of less than 10~\masyr. From this list we then selected stars whose proper motions agreed with the Octans space motion, $(U,V,W)=(-13.7,-4.8,-10.9)$~\kms\ \citep{Murphy15}, at better than 2$\sigma_\mu$, where $\sigma_\mu$ is the total proper motion uncertainty. Using the resulting kinematic distances we also required each candidate lie within the spatial bounds of the Octans/Octans-Near \citep{Zuckerman13} complex, $-100<X<150$~pc, $-150<Y<50$~pc and $-100<Z<50$~pc. To further limit the sample, we finally required a match against near-UV photometry from the \emph{GALEX} mission. Following these cuts, 369 candidates remained over the SPM4 footprint ($\delta<-20$~deg).  After visual inspection to eliminate galaxies and objects with obviously erroneous photometry, spectroscopic observations concentrated on the reddest stars lying above a Pleiades isochrone \citep{Stauffer07} with the best proper motion matches and near-UV excesses \citep{Rodriguez11}. 

We obtained spectroscopic observations of 25 high-priority candidates and a selection of velocity and spectral type standards using the Wide Field Spectrograph \citep[WiFeS;][]{Dopita07} on the Australian National University 2.3-m telescope at Siding Spring Observatory during 2015 March 6--11. The instrument set up, data reduction and analysis, including the derivation of spectral types and radial velocities, was the same as that described in \cite{Murphy15} and we refer the reader to that work for more information. All spectra were flux calibrated using observations of the white dwarf EG~21 \citep{Bessell99}. Further WiFeS observations of two interesting candidates (\starshort{205} and \starshort{91}, see Section\,\ref{sec:scocen}) were taken with the same instrument settings between 2015 March 30 and April 6 as part of another programme\footnote{Reduced spectra of the lithium-rich stars and machine-readable versions of Table A1 can be found on the \textsc{figshare} service at the following DOI:\\ \url{http://dx.doi.org/10.6084/m9.figshare.1491506}}.

\section{Results}

\begin{table*}
\centering
\begin{minipage}{\textwidth}
\caption{Lithium-rich candidates observed with 2.3-m/WiFeS during 2015 March--April. Spectral types are estimated from $V-K_{s}$ colours and comparison to standards. $V$ magnitudes are from SPM4, except for \protect\starshort{78} and \protect\starshort{83} which are from APASS-DR6. Photometry for \protect\starshort{71}, \protect\starshort{82} and \protect\starshort{91} are derived from photographic plates and will be of poorer quality than SPM4 CCD photometry. \protect\starshort{82} has a broadened cross correlation function (CCF) which may indicate binarity (see text).  Candidates displaying emission lines in their spectra are noted (see Table\,\ref{table:halpha}).}
\label{table:lirich}
\begin{tabular}{ccccccccl}
\hline
2MASS designation & Spectral Type & $V$ & $K_{s}$ & EW(Li) & EW(H$\alpha$) & $v_{10}(\text{H}\alpha)$ & RV & Notes\\ 
($J$2000) & ($\pm$0.5) & (mag) & (mag) & ($\pm$50 m\AA) & ($\pm$1 \AA) & ($\pm$10 \kms) &($\pm$2 \kms) \\
\hline
\starlong{74} & M4.9 & 16.72 & 10.70 & 630 & $-$11.5 & 172 & 11.0 & -- \\
\starlong{71} & M4.9 & 16.03 & 10.03 & 600 & $-$10.9 & 162 & 20.0 & TWA 34; [\ion{O}{i}]\\
\starlong{77} & M4.9 & 17.08 & 11.08 & 800 & $-$11.5 & 186 & 19.0 & --\\
\starlong{78} & M5.1 & 15.58 & 9.43 & 700 & $-$11.3 & 151 & 2.0 & --\\
\starlong{82} & M4.7 & 14.55 & 8.72 & 650 & $-$9.9 & 312 & 11.0 & Broad CCF\\
\starlong{83} & M4.8 & 15.77 & 9.85 & 700 & $-$9.5 & 169 & 6.0 & -- \\
\starlong{205} & M5.0 & 17.35 & 11.25 & 650 & [$-$63, $-$27] & [238, 331] & 16.0 & \ion{He}{i}\\
\starlong{91} & M5.0 & 17.31 & 11.27 & 600 & [$-$91, $-$33] & [236, 341] & 9.0 &  \ion{He}{i}, [\ion{O}{i}], \ion{Na}{i}\\
\hline
\end{tabular}
\end{minipage}
\end{table*}

\begin{figure*} 
   \centering
   \includegraphics[width=0.495\linewidth]{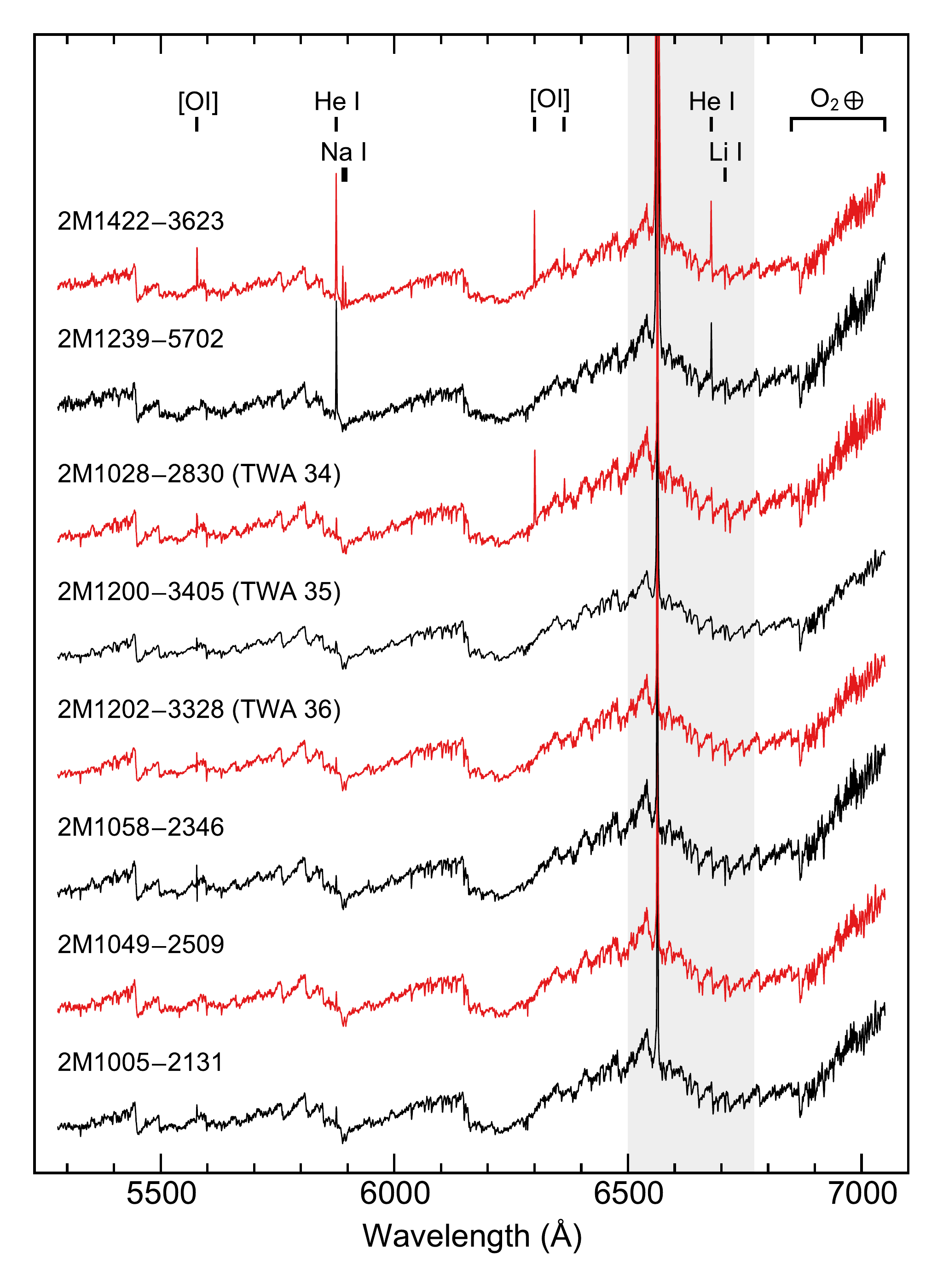} 
   \includegraphics[width=0.495\linewidth]{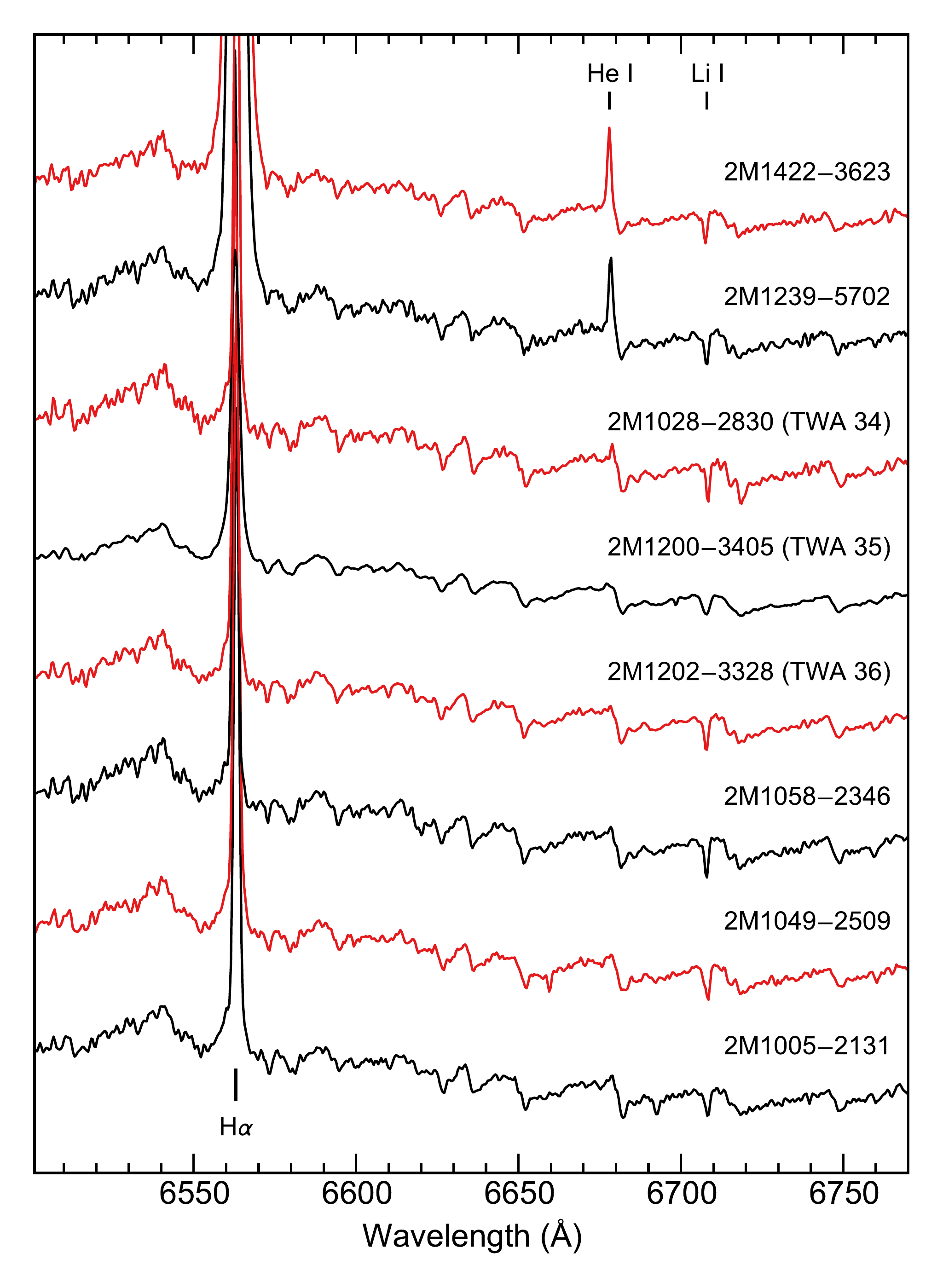} 
   \caption{2.3-m/WiFeS $R7000$ spectra of the lithium-rich stars identified in this work. The full 5300--7050~\AA\ spectra are plotted in the left panel and the shaded region around H$\alpha$ and \ion{Li}{i} $\lambda$6708 is enlarged on the right. All spectra are normalised over 6100--6150~\AA. The broader spectral features of \protect\starshort{82} (especially around \ion{Li}{i}) are readily apparent. Only first epoch spectra of \protect\starshort{91} and \protect\starshort{205} are plotted (see Table\,\ref{table:halpha}). }   \label{fig:spectra}
\end{figure*}

Only 4/25 candidates had radial velocities within 5~\kms\ of those predicted by Octans membership. The relationship of these stars (only one of which is lithium-rich) to Octans will be discussed in a future work. Eight candidates showed strong \ion{Li}{i} $\lambda$6708 absorption ($\text{EW(Li)}\geq 100$~m\AA). These stars are listed in Table~\ref{table:lirich} and their WiFeS $R7000$ spectra are shown in Fig.~\ref{fig:spectra}. The detection of lithium is an indicator of youth in M-type stars, but with a strong stellar mass dependence \citep{Soderblom10}. At ages of 10--20~Myr lithium absorption disappears at early and mid-M spectral types only to reappear at lower masses as one crosses the LDB \citep[e.g.][]{Basri96}. With spectral types of $\sim$M5, the eight stars in Table~\ref{table:lirich} are demonstrably younger than the Pleiades, \citep[LDB M6.5;][]{Stauffer98}, and are likely no older than the nearby $\sim$20~Myr $\beta$~Pictoris Association \citep[LDB M4.5;][]{Binks14}. The sky positions and proper motions of the lithium-rich stars are plotted in Fig.~\ref{fig:skyplot} in Galactic coordinates. All either lie within or near TWA and the adjacent Sco-Cen. 

\subsection{New members of the TW Hydrae association}\label{sec:twa}

\starshort{78}, \starshort{82} and \starshort{83} have positions and proper motions similar to nearby TWA members (Fig.~\ref{fig:skyplot}). The latter two stars lie within 2~deg of the confirmed M1 member TWA~23 \citep{Song03} and all three have strong lithium absorption similar to other M-type members.  \citet{Shkolnik11} identified \starshort{82} as a UV-bright TWA candidate but did  not report any H$\alpha$ emission in their Magellan/MagE spectrum. Based on this non-detection they disregarded \starshort{82} as a possible member and did not obtain a higher-resolution spectrum or radial velocity.  However, subsequent examination of the raw MagE frame shows clear H$\alpha$ emission and that the star was inadvertently excluded as a likely member (E. Shkolnik, personal communication). The WiFeS spectrum of \starshort{82} plotted in Fig.~\ref{fig:spectra} has strong H$\alpha$ emission ($\text{EW} = -9.9$~\AA) typical of an active mid-M dwarf. The broad velocity width ($v_{10}=312$~\kms) may be the result of spectroscopic binarity (see Section\,\ref{sec:binary}). We find \starshort{82} has a \emph{GALEX} $NUV-J$ excess consistent with other nearby young stars \citep{Rodriguez11}, while \starshort{78} and \starshort{83} are borderline UV-excess sources. 

\begin{figure}
   \centering
   \includegraphics[width=0.99\linewidth]{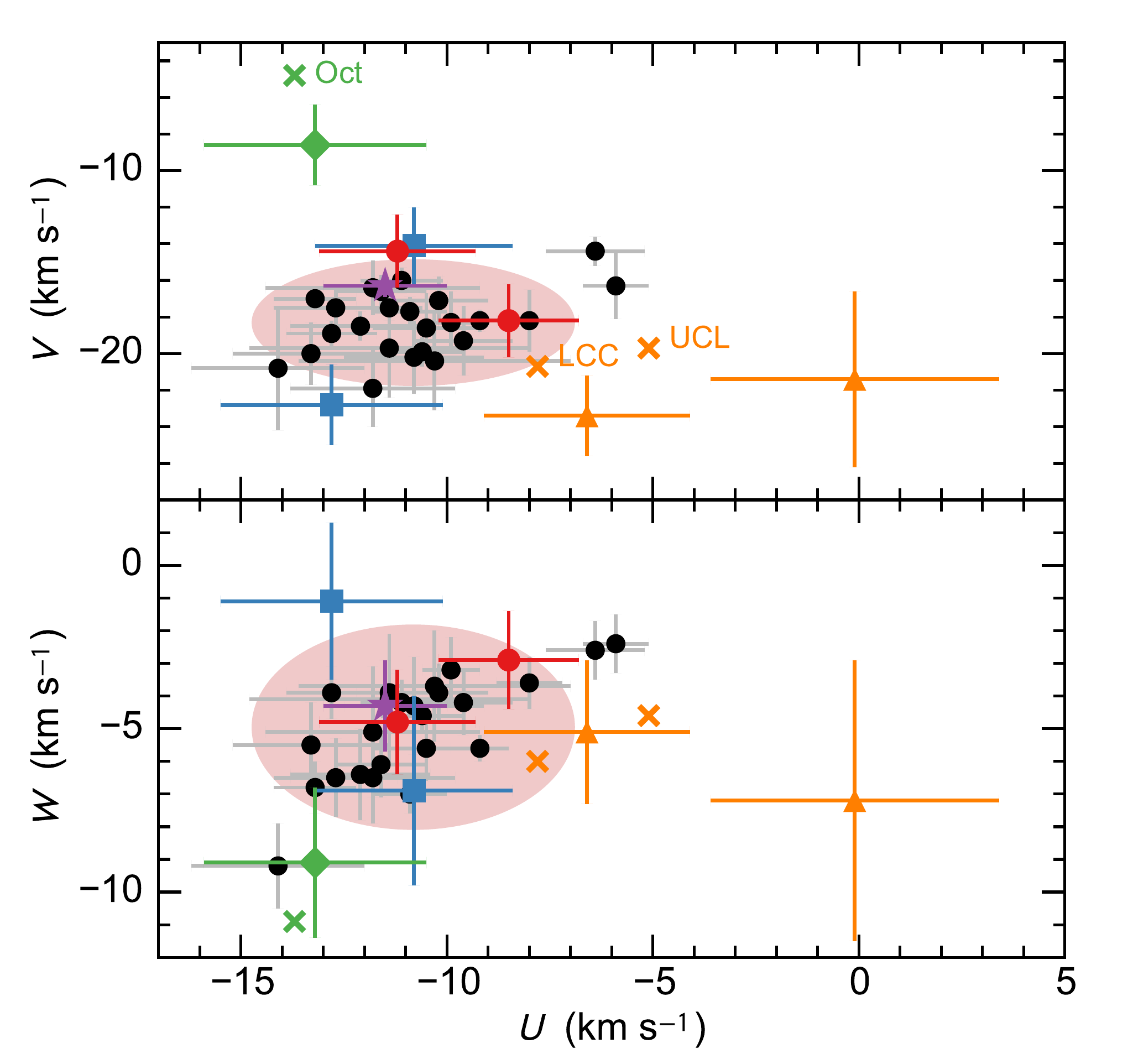} 
   \caption{Candidate space motions (see Table~\ref{table:uvw}), coloured by their proposed memberships: \protect\starshort{82} and \protect\starshort{83} (TWA, red points), \protect\starshort{71}/TWA\,34 (TWA, purple star), \protect\starshort{78} (Octans/TWA?, green diamond), \protect\starshort{205} and \protect\starshort{91} (Sco-Cen, orange triangles) and  \protect\starshort{74} and \protect\starshort{77} (unknown, blue squares). Also plotted are confirmed TWA members with trigonometric parallaxes \citep[black points, with $\pm$2$\sigma$ range shaded;][]{Ducourant14}.}
   \label{fig:uvw}
\end{figure}

To support the membership of \starshort{78}, \starshort{82} and \starshort{83} in TWA we use the \textsc{banyan ii} web service \citep[version 1.3;][]{Malo13,Gagne14}\footnote{\url{http://www.astro.umontreal.ca/~gagne/banyanII.php}}. \textsc{banyan} is a Bayesian analysis tool which compares a star's Galactic position ($XYZ$) and space velocity ($UVW$) to the observed distributions of young groups and the field. No photometric information is used. It~utilises a naive Bayesian classifier to calculate membership probabilities, marginalising over the unknown distance and taking into account the expected populations (prior probabilities) of each group and the field, as well as uncertainties on the observables. Within \textsc{banyan} we adopt the `young' field hypothesis (age $<$1 Gyr) given the clear youth of the lithium-rich stars. Using only their sky positions and SPM4 proper motions, \textsc{banyan} returns TWA membership probabilities of 83, 99 and 98 per cent for \starshort{78}, \starshort{82} and \starshort{83}, respectively, with the balance of membership probabilities going to the field hypothesis.  The predicted radial velocities of $10.3\pm1.4$ and $9.9\pm1.5$~\kms\ for \starshort{82} and \starshort{83} agree well with the WiFeS values in Table\,\ref{table:lirich}, further supporting the stars' membership in TWA\footnote{\starshort{83} was observed with the ESO 2.2-m/FEROS (programme ID 090.C-0200A) on 2013 February 17 as part of the \emph{GALEX} Nearby Young-Star Survey \citep[GALNYSS;][]{Rodriguez11,Rodriguez13}.  The pipeline-reduced spectrum is heavily contaminated by residual sky emission and strong noise spikes. By combining line velocities for H$\alpha$, H$\beta$ and the \ion{Na}{i} $\lambda$8200 absorption doublet we derive a mean radial velocity of $8.3\pm1.7$ (s.d.)~\kms, in good agreement with the WiFeS and \textsc{banyan} values.}. Including these observations in \textsc{banyan} gives membership probabilities of 99 and 97 per cent, with implied distances of $\sim$60~pc for both stars.  The resultant space motions of \starshort{82} and \starshort{83} are listed in Table~\ref{table:uvw} and plotted in Fig.~\ref{fig:uvw} with confirmed TWA members. Their velocities are in excellent agreement with the other members, as expected. The mean space motion of the 24 members in this diagram is $(U,V,W) = (-10.8, -18.3 -5.0)\pm(2.0, 1.7, 1.6)$~\kms\ (1$\sigma$ variation). \textsc{banyan} uses a different set of 18 members in its TWA model, whose principle spatial and velocity axes are not aligned to standard Galactic coordinates \citep[see][]{Gagne14}.  However, both samples have almost identical mean velocities ($\Delta v<1$~\kms). 

\begin{figure} 
   \centering
   \includegraphics[width=0.99\linewidth]{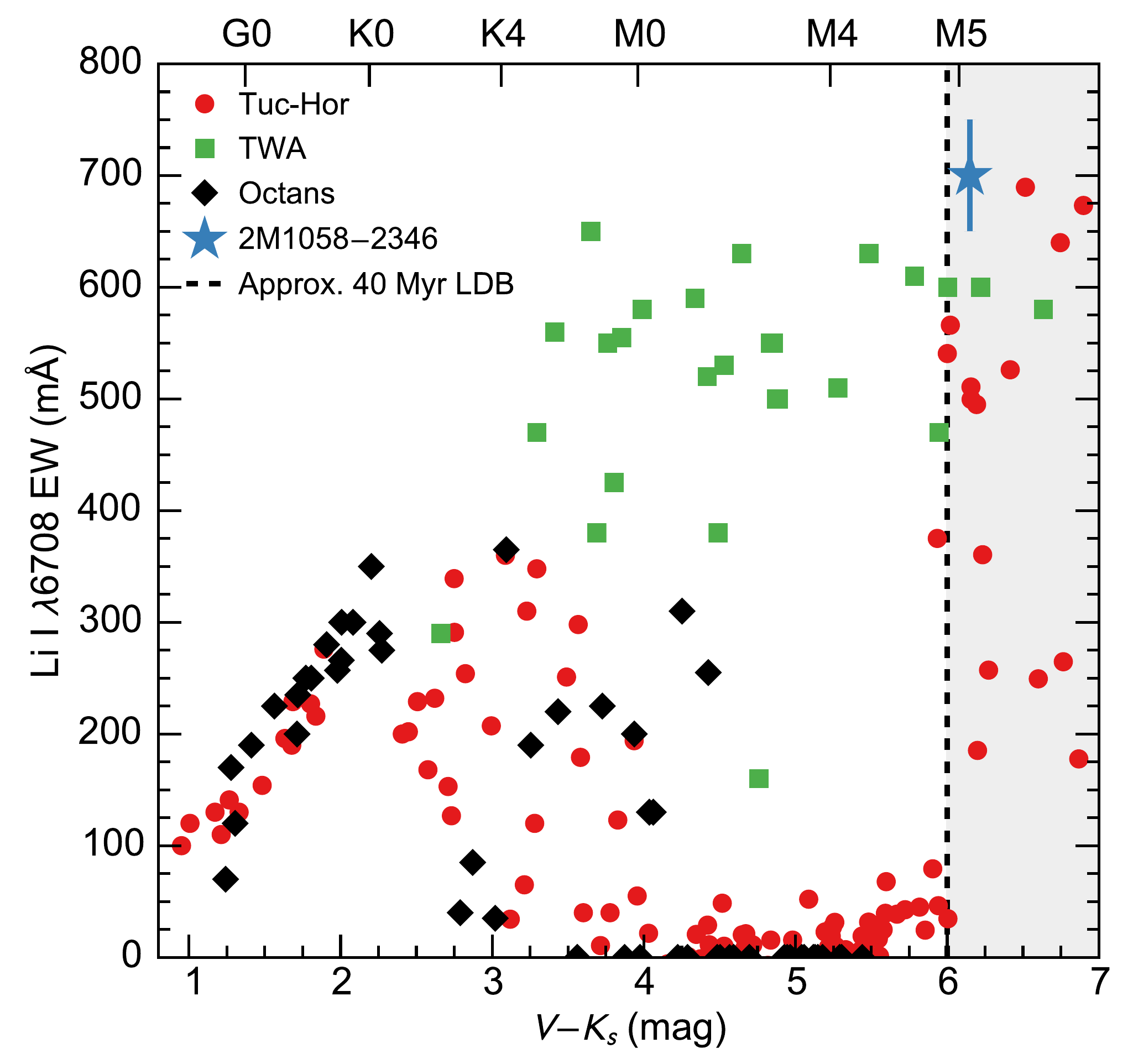} 
   \caption{\ion{Li}{i} $\lambda$6708 EW of \protect\starshort{78} compared to members of TWA \citep{Schneider12b}, Tuc-Hor \citep{da-Silva09,Kraus14} and Octans \citep{Murphy15}. The 30--40~Myr-old Tuc-Hor lithium depletion boundary (LDB) is clearly visible at $V-K_{s}\approx6$ (dashed line).}
   \label{fig:lithium}
\end{figure}

\begin{figure} 
   \centering
   \includegraphics[width=0.99\linewidth]{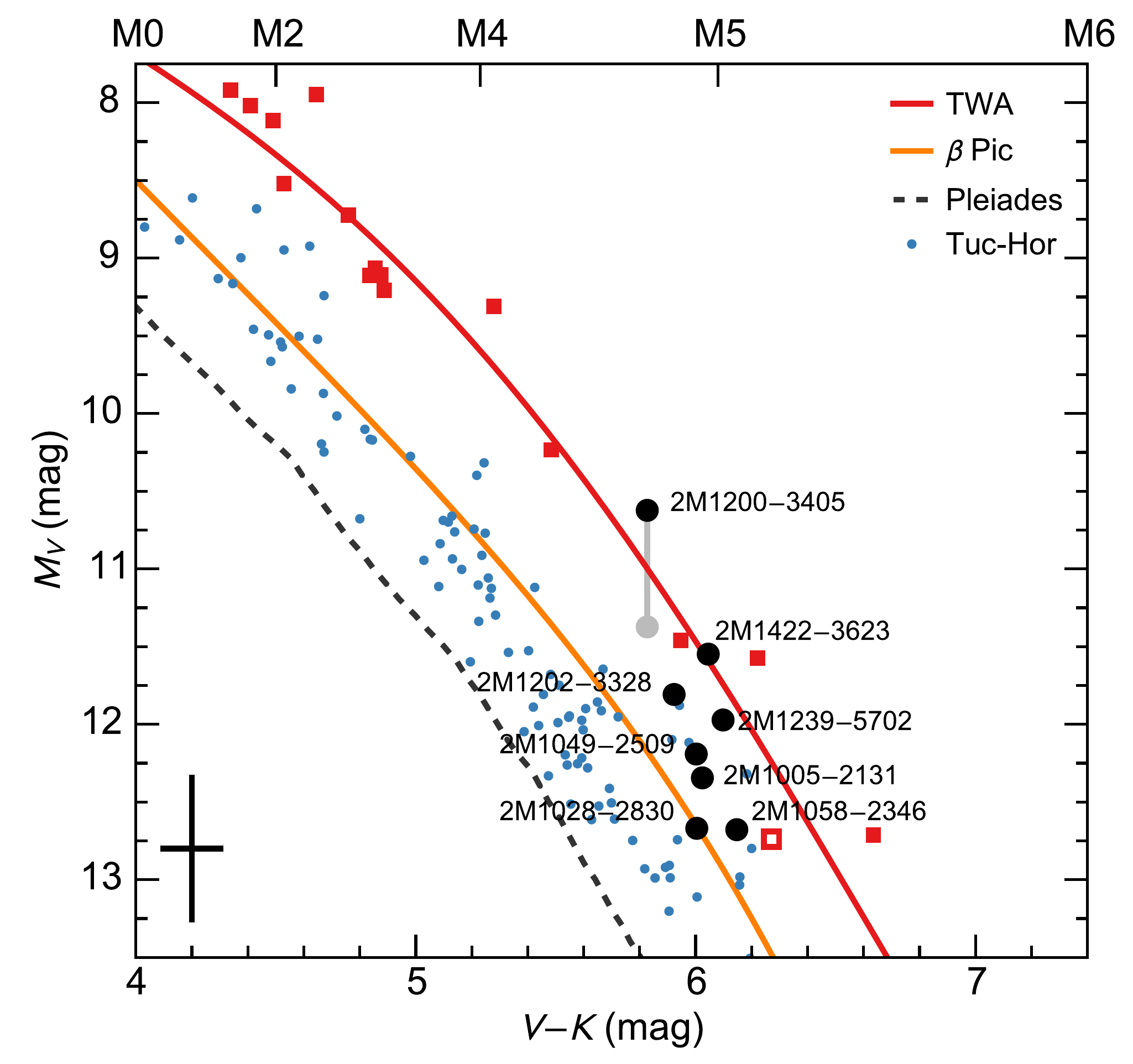} 
   \caption{Candidate colour-magnitude diagram. The grey point assumes \protect\starshort{82} is a corrected equal-brightness system (see text). Also plotted are members of TWA \citep[red squares;][]{Ducourant14} and Tuc-Hor \citep[blue dots;][]{Kraus14}. Lines show fits to TWA (red) and $\beta$~Pic (orange) members from \citet{Riedel11} and the Pleiades sequence \citep{Stauffer07}. The maximal error bar represents a 20~per cent parallax error and 0.1~mag photometric errors.}
   \label{fig:cmd}
\end{figure}

On the contrary, the WiFeS radial velocity  of \starshort{78} ($2\pm2$~\kms) disagrees by 4$\sigma$ with that predicted by \textsc{banyan} for membership in TWA. However, it is the only lithium-rich star to have a space motion consistent with membership in Octans (Fig.~\ref{fig:uvw}; RV$_{\text{Oct}}=-2.2$~\kms)  and the only one associated with a \emph{ROSAT} source \citep[1RXS J105849.9--234623;][]{Voges99}. We estimate $F_{X}\approx2\times10^{-13}$  erg~s$^{-1}$~cm$^{-2}$ (near the \emph{ROSAT} detection limit) and $\log L_X/L_{\text{bol}}\approx-3.2$ using the relations in \citet{Fleming95} and \citet{Pecaut13b}. Detection by \emph{ROSAT} at this spectral type immediately implies the star is closer than the majority of TWA members. \textsc{banyan} predicts a distance of $41\pm4$~pc, similar to that implied by membership in Octans (38~pc), placing it near the periphery of the `Octans-Near' group \citep{Zuckerman13,Murphy15}.  If Octans-Near is the same age as Octans and similar to the Tucana-Horologium Association \citep[30--40~Myr; see discussion in][]{Murphy15}, the strong lithium absorption observed in \starshort{78} ($\text{EW}=700$~m\AA) implies it is on the cool side of the Octans LDB, the first such member to be discovered (see Fig.~\ref{fig:lithium}). Alternatively, given its sky position and proper motion vector it is possible \starshort{78} is a spectroscopic binary member of TWA or that its single-epoch WiFeS radial velocity is erroneous. In this case its lower membership probability in \textsc{banyan} could be due to its location near the edge of the TWA spatial ellipsoid. Further radial velocity measurements are necessary to distinguish between these two scenarios.

\begin{table*}
\centering
\begin{minipage}{\textwidth}
\caption{Kinematics and memberships of the lithium-rich stars. All proper motions are from SPM4. Distances are from \textsc{banyan} (\textsc{ban}) or purely kinematic estimates from the projected group velocity (kin). The latter are assumed to have a 20 per cent parallax error (10 per cent for \protect\starshort{205}). $UVW$ velocities define a right-handed triad with $U$ positive towards the Galactic centre. Uncertainties were calculated following the prescription of \citet{Johnson87} using the uncertainties in proper motion, radial velocity and parallax. The space motion of \protect\starshort{71} (also see Fig.~\ref{fig:uvw}) assumes the higher precision radial velocity ($13.3\pm0.1$~\kms) obtained by Rodriguez et al. from ALMA observations (see text). }
\label{table:uvw}
\begin{tabular}{ccccccccl}
\hline
2MASS designation & $\mu_{\alpha}\cos\delta$ & $\mu_{\delta}$ & Dist. & Ref. & $U$ & $V$ & $W$ & Membership \\ 
($J$2000) & (\masyr) & (\masyr) & (pc) & & (\kms) & (\kms) & (\kms) \\
\hline
\starlong{74} & $-41.7\pm4.6$&$-13.4\pm4.2$ & 75$_{-12}^{+19}$& kin. & $-10.8\pm2.4$&$-14.1\pm2.1$ &$-6.9\pm2.9$ & ?\\
\starlong{71} & $-68.6\pm2.7$ & $-11.4\pm2.5$ & $47\pm6$& \textsc{ban.}& $-11.5\pm1.5$& $-16.3\pm0.6$& $-4.3\pm1.4$& TWA (34)\\
\starlong{77} & $-39.0\pm2.0$ & $-7.9\pm1.8$& 95$_{-16}^{+24}$& kin. & $-12.8\pm2.7$& $-22.8\pm2.2$& $-1.1\pm2.4$& ?\\
\starlong{78} & $-98.3\pm1.9$& $-20.6\pm1.7$& 38$_{-6}^{+9}$ & kin. & $-13.2\pm2.7$&$-8.6\pm2.2$ &$-9.1\pm2.3$ & Oct-Near/TWA?\\
\starlong{82} & $-55.6\pm3.2$ & $-19.2\pm3.0$& $61\pm7$ & \textsc{ban.} & $-8.5\pm1.7$& $-18.2\pm2.0$& $-2.9\pm1.5$ & TWA (35) \\
\starlong{83} & $-58.4\pm3.8$&$-17.3\pm3.5$ & $62\pm7$ & \textsc{ban.} & $-11.2\pm1.9$&$-14.4\pm2.0$ &$-4.8\pm1.6$ & TWA (36) \\
\starlong{205} & $-32.0\pm3.5$& $-10.4\pm3.7$& 119$_{-11}^{+13}$ & kin. & $-6.6\pm2.5$& $-23.4\pm2.2$&$-5.1\pm2.2$ & Sco-Cen (LCC)\\
\starlong{91} &$-17.0\pm6.3$ &$-25.6\pm5.8$ & 142$_{-24}^{+35}$ & kin. & $-0.1\pm3.5$& $-21.4\pm4.8$&$-7.2\pm4.3$ & Sco-Cen (UCL)\\
\hline
\end{tabular}
\end{minipage}
\end{table*}

As a final check, we plot in Fig.~\ref{fig:cmd} the $M_{V}$ versus $V-K_{s}$ colour magnitude diagram (CMD) for the lithium-rich stars, together with empirical sequences for TWA, $\beta$ Pictoris and the Pleiades. \starshort{78} and \starshort{83} have CCD-derived APASS \citep{Henden12} $V$ magnitudes which we adopt over the SPM4 values (but which agree within 0.2~mag), while for \starshort{71}, \starshort{82} and \starshort{91} the SPM4 $V$ magnitude was measured from photographic plates and should be treated with caution. No APASS data exist for these stars. Both \starshort{82} and \starshort{83} fall close to the mean locus of TWA members \citep{Riedel11}. In conjunction with their kinematic and spatial match from \textsc{banyan}, the CMD placement, strong lithium absorption and NUV excesses of \starshort{82} and \starshort{83} confirm their youth and membership in TWA. Following the nomenclature for TWA members, we designate \starshort{82} as TWA 35 and \starshort{83} as TWA 36. Neither star displays excess emission above photospheric levels in 3--22~\micron\ All\emph{WISE} photometry \citep{Wright10}. The CMD position of \starshort{78} is consistent with a $\beta$~Pic-like age or the upper envelope of Tuc-Hor members, but still marginally consistent with TWA. Until a better velocity measurement is available we refrain from assigning the star to either TWA or Octans.

\subsubsection{Possible binarity of \protect\starlong{82}}\label{sec:binary}

The mean width of the cross correlation function (CCF) for \starshort{82} ($\text{FWHM}=3.4$~px) is much broader than the other candidates and standards \citep[$\text{FWHM}<2.5$~px; see fig.~4 of][]{Murphy15}. This broadening is also visible in the raw spectrum (Fig.~\ref{fig:spectra}) and H$\alpha$ $v_{10}$ velocity width (Table\,\ref{table:lirich}), and is usually indicative of fast rotation \citep[e.g.][]{Soderblom89}, unresolved spectroscopic binarity or a combination of the two.  Using \starshort{83} as a narrow-lined template we can reproduce the level of broadening observed in \starshort{82} with either a \vsini\ or SB2 velocity shift of $\sim$50~\kms, close to the $c\Delta\lambda/\lambda \approx 45$~\kms\ WiFeS velocity resolution. No broadening is evident at velocities $\lesssim$20~\kms\ ($c\Delta\lambda/2\lambda$). Such a high implied rotation speed is unusual for single TWA stars \citep{Jayawardhana06} and so may instead arise in the blending of lines from a close companion. However, in the absence of higher resolution observations we cannot yet confirm this. If \starshort{82} is a binary then the unresolved CCF peak we measured should still trace the systemic velocity, assuming the binary is close to equal mass.

\subsubsection{\protect\starlong{71} (=TWA 34)}

Another candidate,  \starshort{71}, is the recently-proposed TWA member and disc host TWA~34 \citep{Schneider12}. We confirm the M4.9 spectral type and $\text{EW(H$\alpha$)}=-9.6$~\AA\ reported by \citeauthor{Schneider12} and measure $\text{EW(Li)}=600$~m\AA\ and $\text{RV}=20\pm2$~\kms\ from our higher-resolution WiFeS spectrum. From the modest strength and velocity width of its H$\alpha$ line ($v_{10} = 162$~\kms), TWA~34 does not appear to be accreting from its disc. However, we also detect strong forbidden [\ion{O}{i}] $\lambda$6300 ($\text{EW}=-3.0$~\AA) and $\lambda$6363 ($-$0.6~\AA) emission, indicating the presence of low-density gas usually attributed to a wind or outflow \citep[e.g.][]{Appenzeller89,Rigliaco13,Zuckerman14}. Due to uncertainties in subtracting the strong auroral sky line, we do not unambiguously detect [\ion{O}{i}] $\lambda$5577  emission from any candidate.  The SPM4 proper motion of TWA~34 yields a \textsc{banyan~ii} membership probability of 95~per cent, with an implied distance and radial velocity of $47\pm6$~pc and $13.5\pm1.3$~\kms.  Rodriguez et al. (2015, in preparation) have obtained ALMA observations of TWA~34 and from a Keplerian fit to the CO velocity map derived a barycentric velocity of $13.3\pm0.1$~\kms, in excellent agreement with the \textsc{banyan} prediction and other TWA members (Fig.~\ref{fig:uvw}). This suggests that our WiFeS radial velocity may be erroneous, although we cannot yet rule out spectroscopic binarity.

\begin{figure*}
   \centering
   \includegraphics[width=0.99\textwidth]{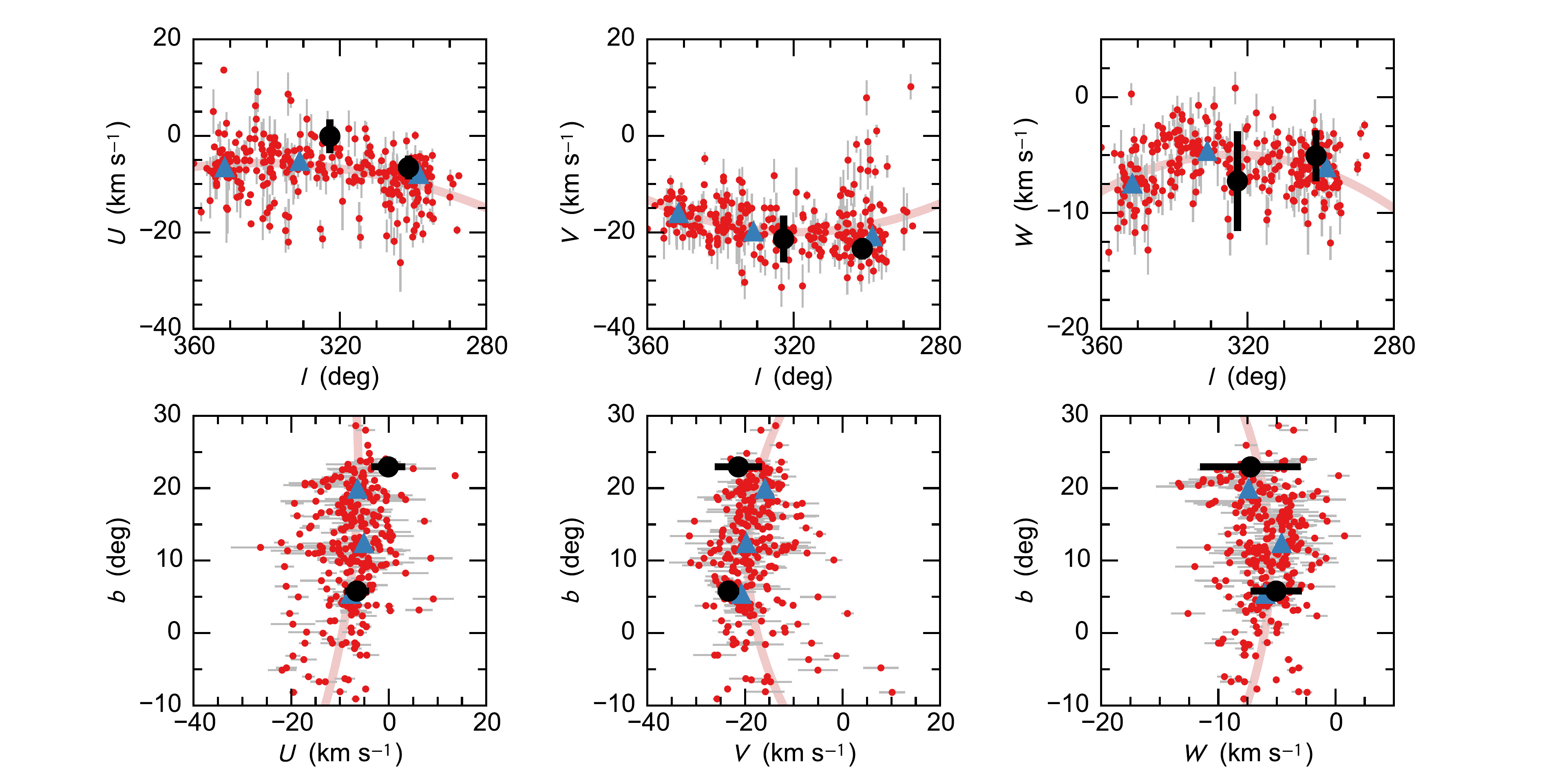} 
   \caption{Space motions of Sco-Cen members from \protect\citet{de-Zeeuw99} (red points) versus Galactic longitude (top) and latitude (bottom). Quadratic fits are shown to guide the eye. The mean velocities of the US, UCL and LCC subgroups (from left to right and top to bottom) are given by the blue triangles, as derived by \protect\citet{Chen11}. The space motions of \protect\starshort{205} and \protect\starshort{91} from Table\,\ref{table:uvw} are plotted as large black points.}
   \label{fig:trends}
\end{figure*}

At 47~pc \starshort{71} is significantly under-luminous compared to other TWA members in Fig.~\ref{fig:cmd}, even considering reasonable errors on its photographic $V$ magnitude. A distance of $\sim$80~pc is required to bring it onto the TWA sequence of \citet{Riedel11}. \citet{Schneider12} found congruent kinematic and photometric distances of $\sim$50~pc when considering $M_{K}$ and an empirical TWA isochrone, but the star is also under-luminous in that band. Its optical and near-infrared spectral energy distribution is well fit by a 3000~K model atmosphere \citep{Bayo08}, consistent with its $V-K_{s}$ colour and spectral type. A possible explanation for the underluminosity is that the photospheric light is intercepted by a flared disc seen at high inclination \citep[similar to $\epsilon$ Cha 11;][]{Fang13}. \citet{Liu15} obtained \emph{Herschel}/PACS photometry of \starshort{71}/TWA\,34 and modelled its disc. Although the best fit inclination angle of $i=60_{-45}^{+7.5}$~deg is not strongly constrained by the data, their models rule out high inclinations ($i\gtrsim80$~deg).

\subsection{Two new accreting M-dwarfs in Sco-Cen}\label{sec:scocen}

The only two other stars in our sample with All\emph{WISE} infrared excesses, \starshort{205} and \starshort{91}, lie within the classical boundaries of the LCC and and UCL subgroups of Sco-Cen, respectively (Fig.~\ref{fig:skyplot}).  Neither star is a member of the seven nearby moving groups tested by \textsc{banyan} (TWA, $\beta$~Pic, Argus, Tuc-Hor, Carina, Columba, AB Dor). The SPM4 proper motions and errors of \starshort{205} and \starshort{91} correspond to kinematic distances of 119$_{-11}^{+13}$~pc and 142$_{-24}^{+35}$~pc, assuming the mean LCC and UCL space motions of \citet{Chen11}. These are in excellent agreement with the 120 and 140~pc mean distances found by \citet{de-Zeeuw99} for early and solar-type LCC and UCL members with \emph{Hipparcos} parallaxes (also see Fig.~\ref{fig:rvcat}). The CMD positions of \starshort{205} and \starshort{91} in Fig.~\ref{fig:cmd} suggest an age of approximately 10 Myr, similar to TWA and consistent with recent age ranges for UCL and LCC \citep{Song12,Pecaut12}.  The age of Sco-Cen and possible age spreads in the association are discussed in greater detail in Section~\ref{sec:scocenage}.

At these distances \starshort{205} and \starshort{91} are 3~\kms\ and 6~\kms, respectively, from the \citeauthor{Chen11} mean space motions (Fig.~\ref{fig:uvw}), compared to uncertainties of 2--4~\kms\ in each velocity component.  However, unlike smaller, more coherent moving groups, OB associations are expected to show spatial and kinematic substructure. To examine the trend and spread of velocities across Sco-Cen, we plot in Fig.~\ref{fig:trends} over 270 high-probability members from \citet{de-Zeeuw99} with radial velocities and updated \emph{Hipparcos} astrometry. This is the largest available compilation of Sco-Cen members with six-dimensional phase space information  and a full description of its construction is given in Appendix~\ref{sec:rvcat}.  The space motions of \starshort{205} and \starshort{91} are perfectly consistent with the surrounding higher-mass members in this diagram.  As expected, the mean subgroup velocities closely follow the longitude and latitude trends, albeit with significant scatter in individual velocities and minimal differentiation into the classical subgroups.  Much of this scatter likely results from observational errors and single-epoch velocities of unknown spectroscopic binaries. For reference, \citet{de-Bruijne99a} found a one dimensional internal velocity dispersion of $<$1--1.5~\kms\ from \emph{Hipparcos} astrometry. Figs.~\ref{fig:trends} and \ref{fig:rvcat} appear to confirm the suggestion of \citet{Rizzuto11} that Sco-Cen can be well modelled as continua in spatial position and velocity rather than three discrete subgroups. The addition of many more low mass members with radial velocities and \emph{Gaia} parallaxes will better resolve the velocity and spatial substructure of Sco-Cen in coming years. 

\begin{figure*} 
   \centering
   \includegraphics[width=0.495\linewidth]{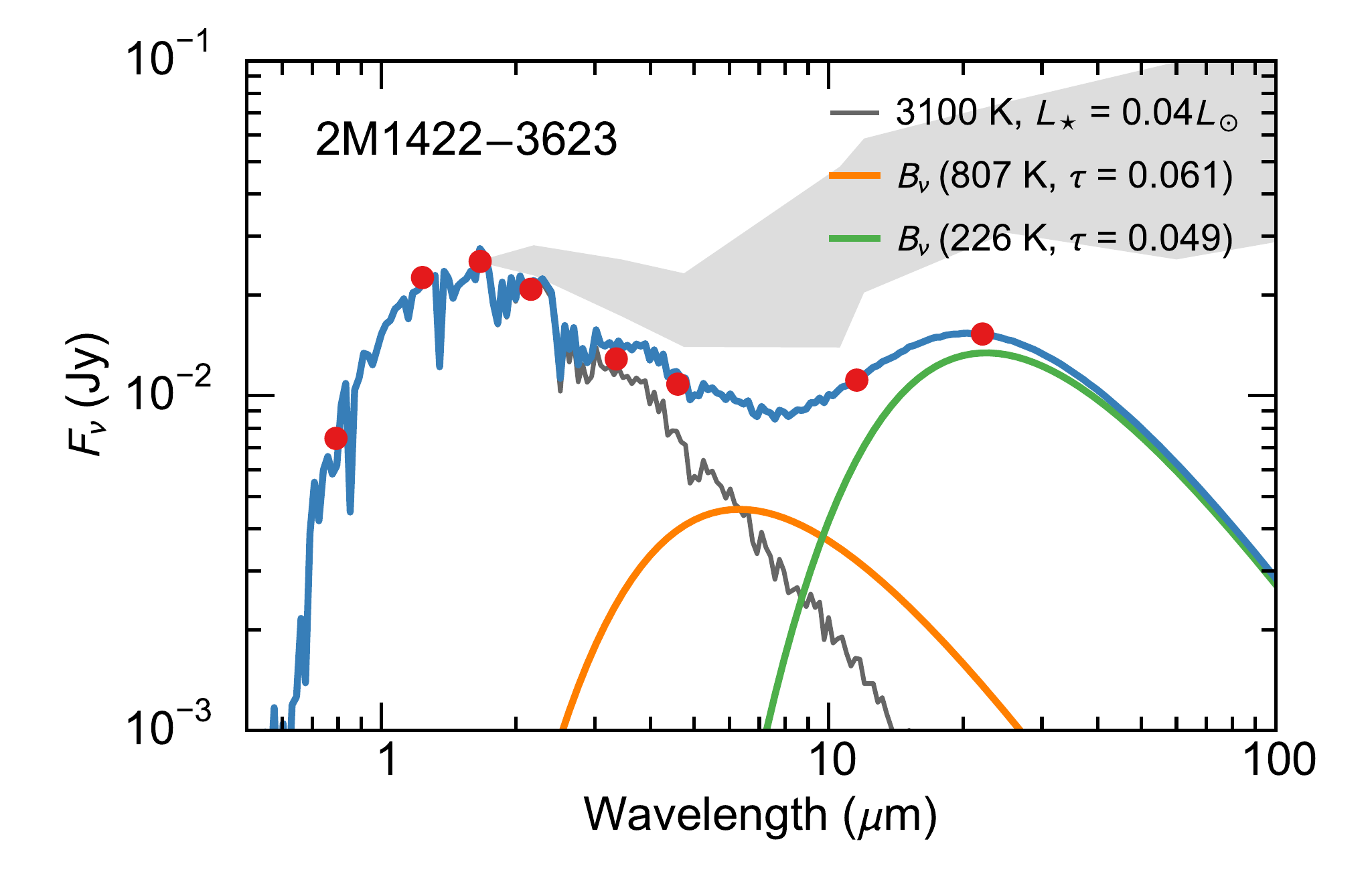}
      \includegraphics[width=0.495\linewidth]{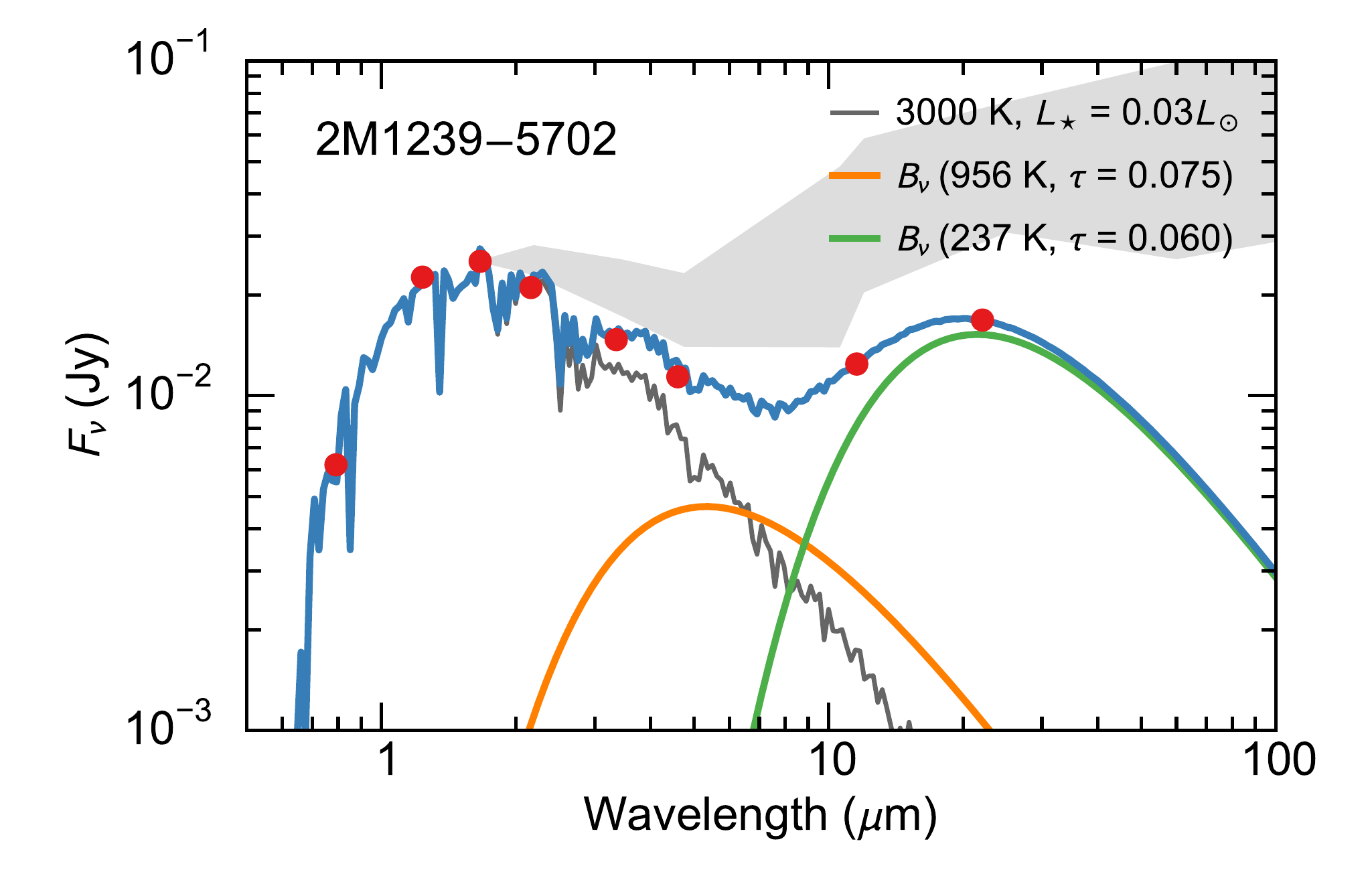} 
   \caption{Spectral energy distributions (SEDs) of \protect\starshort{91} (left) and \protect\starshort{205} (right). Photometric measurements from DENIS ($I$), 2MASS and \emph{WISE} are plotted in red. The total SED (blue line) is approximated by a solar metallicity BT-Settl model \citep[grey line;][]{Allard12} fitted to the DENIS and 2MASS data, and two blackbodies (green and orange lines).  The characteristic emitting area of each component is given by the factor $\tau=L_{\text{IR}}/L_{\star}$. The interquartile range of K5--M2 Class I Taurus sources from \protect\citet{DAlessio99} is shown by the shaded region, normalised at 1.6~\micron.}
   \label{fig:sed}
\end{figure*}

The few Myr-old Lupus star forming region \citep{Comeron08} lies within UCL and its members have similar distances and kinematics to Sco-Cen. It is therefore plausible that \starshort{91} may be a young, outlying member of Lupus instead of the older Sco-Cen population. However, we do not believe this is likely. \starshort{91} is some 15--20~deg from the dark clouds which define the star forming region, and a recent kinematic and membership reanalysis of stars in the area by \cite{Galli13} did not find any legitimate `off-cloud' Lupus members within 9~deg of \starshort{91}. The results of  \citet{Galli13} strongly support the hypothesis that the majority of the $\sim$150 late-type, older pre-main sequence stars found around the Lupus clouds in X-ray surveys \citep[e.g.][]{Krautter97,Wichmann97} are in fact members of UCL/Sco-Cen \citep{Mamajek02,Preibisch08}.  \starshort{205} is situated close to the centre of LCC, only 1 deg from the foreground M-giant and Southern Cross member $\gamma$ Cru. It lies a few degrees north of the so-called `Park \& Finley' stars \citep{Park96,Preibisch08}, a group of four young K and M-type LCC members discovered in the vicinity of the B0.5 member $\beta$ Cru \citep{Feigelson97,Alcala02}. None of these stars appear to be accreting or host circumstellar discs.

\subsubsection{Disc and accretion properties}

The spectral energy distributions of \starshort{91} and \starshort{205} are compiled in Fig.~\ref{fig:sed}. Both stars exhibit strong excesses in the $W3$ (12~\micron) and $W4$ (22~\micron) \emph{WISE} bands. These are consistent with evolved, `homologously depleted' discs which are becoming optically thin without forming an inner hole or gap, and show a strong flux decrement at all infrared wavelengths  relative to optically-thick `full' discs  \citep{Espaillat12,Luhman12b}.  The observed $(K_{s}-W4)$ excesses of \starshort{91} and \starshort{205} are similar to discs around K5--M5 stars in Upper Scorpius (age $\sim$10~Myr), which already show evidence of significant evolution compared to Taurus \citep{Luhman12b}. Although not necessarily physical, double-temperature blackbody models with a cool $\sim$230~K outer component and a warmer 800--900~K inner component  provide a good fit to the \emph{WISE} photometry in both cases. Assuming $L_{*} = 0.03\,L_{\sun}$ (Fig.\,\ref{fig:sed}) and large (blackbody) grains, these temperatures correspond to disc radii of 0.25~au and $\sim$0.02~au, respectively \citep{Backman93}. Further infrared and millimetre observations (e.g. with VLT/VISIR and ALMA) will help elucidate the structure and chemistry of the discs around \starshort{91} and \starshort{205}, for instance by resolving the 10~\micron\ silicate feature if it exists, or detecting cold dust and CO gas emission.

\begin{figure} 
   \centering
   \includegraphics[width=0.99\linewidth]{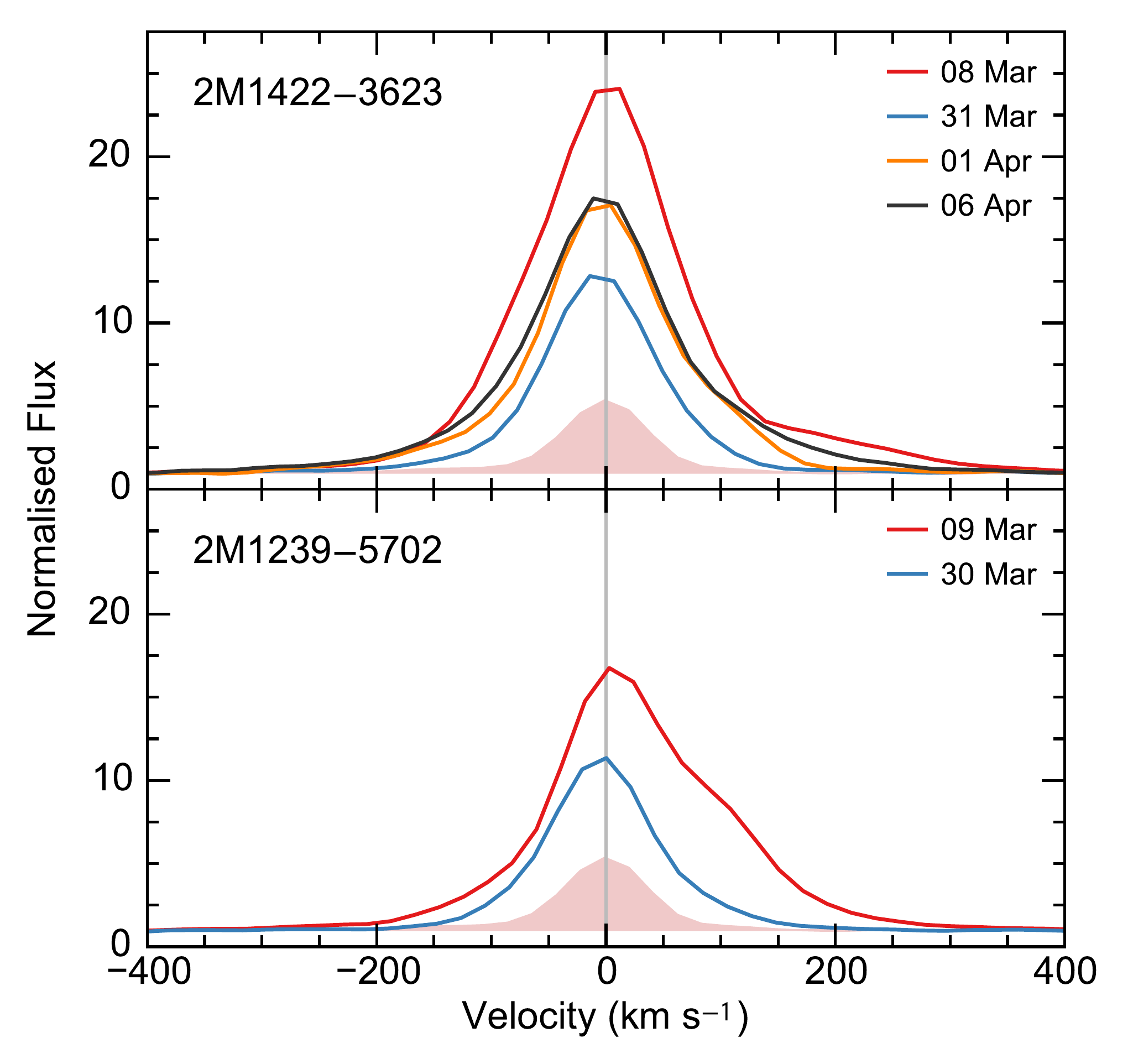} 
      \caption{2.3-m/WiFeS RV-corrected H$\alpha$ velocity profiles of the accretors \protect\starshort{91} (top) and \protect\starshort{205} (bottom). For comparison, each panel also shows the profile of the new TWA member and non-accretor \protect\starshort{83} (shaded region; $\text{EW}=-9.5$~\AA, $v_{10}=169$~\kms).}
   \label{fig:halpha}
\end{figure}

\begin{table*}
\centering
\begin{minipage}{\textwidth}
\caption{Emission line equivalent widths in \protect\starshort{205} and \protect\starshort{91}. The velocity width of the H$\alpha$ line at 10 per cent of peak flux ($v_{10}$) is also given. Estimated uncertainties are $\pm$10~\kms\ for $v_{10}$,  $\pm$0.1~\AA\ for all lines other than H$\alpha$ and $\pm$2~\AA\ for the broader H$\alpha$ line.}
\label{table:halpha}
\begin{tabular}{llcccccccc}
\hline
Star & UT Date &  $v_{10}(\text{H}\alpha)$ & H$\alpha$ & \ion{He}{i} $\lambda$5876 & \ion{He}{i} $\lambda$6678 & [\ion{O}{i}] $\lambda$6300 & [\ion{O}{i}] $\lambda$6363 & \ion{Na}{i} $\lambda$5896 & \ion{Na}{i} $\lambda$5890\\
& & (\kms) & (\AA) &  (\AA) & (\AA) & (\AA) & (\AA) & (\AA) & (\AA)\\
\hline
 \starshort{205} & 2015 March 9 & 331 & $-$63 & $-$5.3 & $-$1.2 & \dots & \dots & $-$0.2 & $-$0.2\\
& 2015 March 30  & 238 & $-$27 & $-$2.4 & $-$0.4 & $-$0.4 & \dots & $-$0.6 & $-$0.9 \\
\hline
\starshort{91} & 2015 March 8 & 335 & $-$91 & $-$5.7 & $-$1.3 & $-$2.5 & $-$0.5 & $-$3.2 & $-$1.9\\
& 2015 March 31 & 236 & $-$33 & $-$3.4 & $-$0.6 & $-$2.8 & $-$0.8 & $-$2.0 & $-$1.6\\
& 2015 April 1 & 303 & $-$54 & $-$1.7 & $-$0.3 & $-$3.7 & $-$1.0 & $-$1.8 & $-$1.0\\
& 2015 April 6 &  341 & $-$68 & $-$3.1 & $-$0.9 & $-$2.4 & $-$0.6 & $-$2.4 & $-$0.8\\
\hline
\end{tabular}
\end{minipage}
\end{table*}

Accretion of gas from the inner disc onto the star is usually accompanied by enhanced Balmer and other line emission \citep{Muzerolle98b}. H$\alpha$ velocity profiles for \starshort{205} and \starshort{91} are given in Fig.~\ref{fig:halpha}.  The broad and strong emission at all epochs (also see Table\,\ref{table:halpha}) confirms both stars are accreting, with EWs greater than the 15--18~\AA\ threshold for M3--M6 stars suggested by \citet{Fang09} and $v_{10}$ velocity widths in excess of the 270~\kms\ accretion criterion of \cite{White03}\footnote{\citet{Jayawardhana03} suggest a lower accretion threshold of 200~\kms\ for M5--M8 objects. Whatever the adopted criterion, it is clear from the large EWs and the presence of other indicators  that \starshort{205} and \starshort{91} are accreting from gas-rich inner discs.}.  Both stars presented weaker H$\alpha$ emission in the second-epoch observations, falling just below the 270~\kms\ threshold but still above the EW limit. Subsequent observations of \starshort{91} show a recovery in both EW and $v_{10}$. Applying the $v_{10}(\text{H}\alpha)$-$\dot{M}$ relation of \citet{Natta04} to the velocities in Table~\ref{table:halpha} yields mass accretion rates of $\sim$$10^{-9.5}$~\msun\ yr$^{-1}$ at peak and a factor of ten lower at minima, comparable to those in older groups like TWA and $\eta$ Cha \citep{Muzerolle00,Lawson04}. Similar behaviour is seen in the $\sim$6~Myr-old $\eta$ Cha members RECX~5 and 2M0820--8003, which are borderline accretors but have undergone sporadic episodes of strong accretion in multi-epoch observations \citep{Lawson04,Murphy11}. Unlike RECX~5, which hosts a transitional disc with an inner hole and gap \citep{Bouwman10}, 2M0820--8003, \starshort{205} and \starshort{91} show excesses at $\lesssim$5~\micron, indicating the presence of dust in the inner disc. Their variable accretion is presumably driven by  `clumpy' residual gas-rich material in this region. 

In addition to H$\alpha$ we also observed \ion{Na}{i} D and forbidden [\ion{O}{i}] $\lambda$6300/6363 emission in \starshort{91} and strong \ion{He}{i} $\lambda$5876/6678 emission in both stars (Fig.~\ref{fig:spectra} and Table\,\ref{table:halpha}). The strength of the \ion{Na}{i} and \ion{He}{i} emission was generally correlated with H$\alpha$, as expected from their common origin in magnetospheric infall regions \citep{Muzerolle98b}, whereas the outflow-driven forbidden emission remained approximately constant. Continuum excess is another indicator of accretion and activity in T Tauri stars, particularly in the UV \citep{Bertout88}. \starshort{205} and \starshort{91} both show enhanced \emph{GALEX} near-UV emission compared to the other lithium-rich stars and most stars in the \cite{Torres08} sample of young moving group members.  Comparing  \ion{Li}{i} $\lambda$6708 line strengths we deduce that only the 2015 March 8 spectrum of \starshort{91} was noticeably veiled (when accretion was strongest). On that night we measure an EW of 550~m\AA, compared to values of 600--630~m\AA\ at the other three epochs.

\subsection{Other lithium-rich stars}

\starshort{74} and \starshort{77} both have smaller proper motions, inconsistent with neighbouring TWA members and indicative of larger distances. We find satisfactory matches to the TWA space motion at 75 and 95~pc, respectively (Fig.~\ref{fig:uvw}). However, at these distances the stars are very far from other TWA members in space \citep[e.g.][]{Weinberger13} and still sit below the mean CMD trend in Fig.~\ref{fig:cmd}.  Adopting the UCAC4 proper motion for \starshort{74}  ($-55.3\pm7.1$,  $-9.7\pm5.3$~\masyr) increases its \textsc{banyan} membership probability from 13 to 50~per cent (including RV), but at the implied distance of 42~pc the star lies near the Pleiades sequence and its predicted proper motion is much larger than observed.

\starshort{77} was observed with VLT/UVES on 2014 May~11 (programme ID 093.C-0133A) as part of GALNYSS. Examining the pipeline-reduced archival spectrum we find $\text{EW(Li)}=680\pm30$~m\AA, $\text{EW(H$\alpha$)}=-22\pm3$~\AA, and $-300$~m\AA\ \ion{Ca}{ii} infrared triplet emission. The radial velocity of the \ion{Na}{i} $\lambda$8200 absorption doublet ($17\pm1$~\kms) agrees with our lower-resolution WiFeS measurement. The star was also serendipitously observed by both the \emph{Chandra} and \emph{XMM-Newton} X-ray satellites. Its 0.2--12~keV \emph{XMM} luminosity at 95~pc   \citep[10$^{28.9}$~erg~s$^{-1}$;][]{Rosen15} and $\log L_{X}/L_{\text{bol}}\approx-3.0$ are both consistent with an age less than the Pleiades.   Both \starshort{74} and \starshort{77} are clearly young, but their exact membership awaits further observations, particularly parallaxes. 

\section{Discussion}

\subsection{Other accretors in Sco-Cen}\label{sec:otheraccretors}

Outside of Upper Scorpius, \starshort{205} and \starshort{91} join only a handful of known accretors in the older UCL and LCC subgroups. \citet{Shkolnik11} suggested the M3.5 LCC member 2MASS J11311482--4826279 was accreting based on its broad H$\alpha$ line ($v_{10}=233$~\kms). However, given its weak EW ($-7.3$~\AA) and that the star has no excess emission visible in \emph{WISE}, we suggest the broad H$\alpha$ line \citeauthor{Shkolnik11} observed was the result of chromospheric activity. Flares can give rise to broad, multi-component H$\alpha$ emission like that seen in 2M1131--4826 and are usually characterised by lower EWs than observed in accretors \citep{Montes98,Jayawardhana06,Murphy11}. \citet{Rodriguez11} proposed another M3.5 LCC accretor, 2MASS J13373839--4736297, based on a strong ($-13.7$~\AA) and asymmetric H$\alpha$ line. \citet{Schneider12} confirmed the star hosts a disc with excesses in all four \emph{WISE} bands. Accretion thus seems a likely explanation for the observed line emission. \citet{Zuckerman15} report 2M1337--4736 is a 10~arcsec wide binary of near equal brightness. The secondary does not appear to possess a disc. 

The three confirmed M-type accretors in Sco-Cen (2M1337--4736A, \starshort{205} and \starshort{91}) are joined by the Herbig Ae star HD~139614 and two accreting F stars; AK Sco and HD~135344 \citep{Preibisch08}. These authors noted HD~139614 and 135344 lie near the western edge of the Lupus clouds and so may be younger than the median age of UCL \citep[16~Myr;][]{Pecaut12}. \citet{Mamajek02} proposed the K1 T Tauri star MP Mus (=PDS 66) as a southern member of LCC. It was the only star which showed signs of accretion ($\text{EW(H$\alpha$)}=-39$~\AA) in their survey of 110 G and K-type UCL and LCC stars. However, \citet{Murphy13} have disputed the membership of MP~Mus in LCC, claiming a better kinematic and distance match to the 3--5~Myr $\epsilon$ Chamaeleontis association which overlaps the south of LCC.

In light of the increasing number of accretors in Sco-Cen, membership of the inconclusive TWA member and confirmed accretor 2MASS J12071089--3230537 \cite[TWA~31;][]{Shkolnik11,Schneider12b} deserves reevaluation. TWA~31 was rejected by \cite{Ducourant14} on the basis of its poor USNO-B1.0 proper motion. \textsc{banyan} gives a membership probability of only 28~per cent at $59\pm7$~pc. Its SPM4 proper motion, $(-52\pm8,-18\pm8)$~\masyr, is a better match to TWA, with a membership probability of 98~per cent at $61\pm7$~pc. However, with $V_{\text{SPM4}}=18.45$ (photographic) TWA~31 would be under-luminous at this distance. It is not found in UCAC catalogues and its PPMXL proper motion agrees with USNO-B1.0. At its proposed photometric distance of 110~pc \citep{Shkolnik11}, TWA 31 has an SPM4 space motion somewhat similar to other TWA members and a reasonable CMD position, but is much further away than the majority of TWA stars. Only TWA 15AB is more distant ($d\approx115$~pc), but it is far enough south to conceivably be an LCC member \citep[but see discussion in][]{Ducourant14}. Until better astrometry are available the final status of TWA~31 remains unclear. 

\subsection{Implications for the age of Sco-Cen and disc lifetimes}\label{sec:scocenage}

\begin{figure} 
   \centering
   \includegraphics[width=0.99\linewidth]{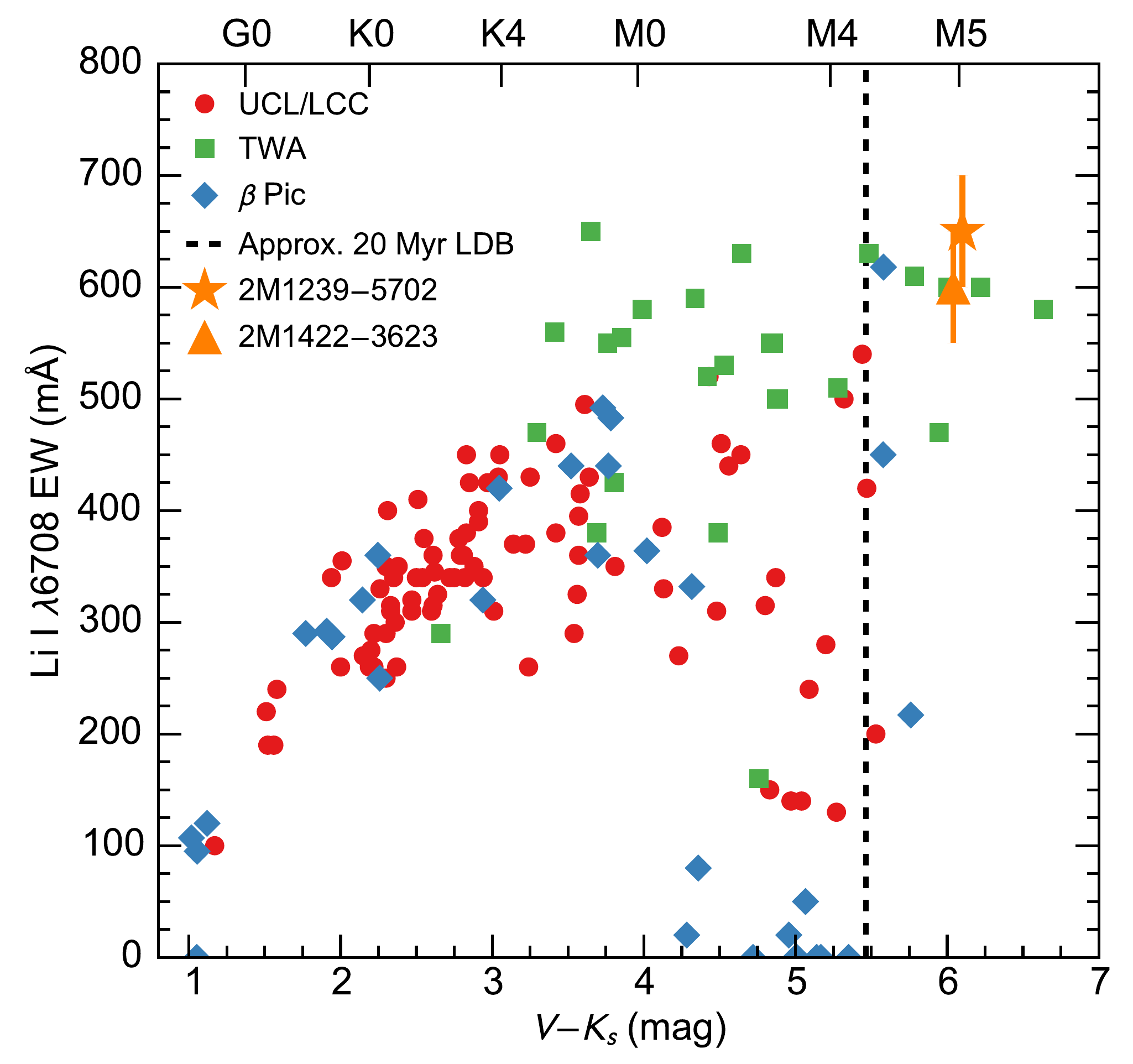} 
   \caption{\ion{Li}{i} $\lambda$6708 EWs of \protect\starshort{205} (star) and \protect\starshort{91} (triangle) compared to new UCL and LCC members from \citet{Song12}. Members of TWA \citep{Schneider12b} and $\beta$~Pic \citep{da-Silva09,Binks14} are plotted for comparison. The $\sim$20 Myr-old $\beta$ Pic LDB is also given (dashed line). Sco-Cen members have lithium depletion levels between TWA and $\beta$ Pic, corresponding to an age of 10--20~Myr.}
   \label{fig:lithium_scocen}
\end{figure}

The fraction of accreting stars in young groups and star forming regions is observed to fall rapidly with age, with $e$-folding time-scales of 2--3~Myr \citep{Mamajek09,Fedele10}.  For instance, \citet{Frasca15} recently found accretor fractions of 35--40~per cent for the $\sim$2~Myr-old Cha I population, falling to only 2--4~per cent in the older ($\geq$10~Myr) $\gamma$ Velorum cluster. Similar behaviour is observed in young moving groups, where the accretor fraction in the $<$10~Myr $\eta$ and $\epsilon$ Cha groups is $\sim$30~per cent \citep{Murphy13} but by the age of Upper Scorpius (10 Myr) has fallen to $\sim$10~per cent \citep{Lodieu11} and in UCL and LCC (16--17~Myr) is at most a few per cent \citep{Mamajek02}. Complicating matters are recent recalibrations of pre-main sequence cluster and association ages by \citet{Bell13}, which have increased the ages of some benchmark groups by up to a factor of two and therefore increased average circumstellar disc lifetimes. Nevertheless, if \starshort{205} and \starshort{91} are at the median age of LCC and UCL then they are incredibly rare objects. Assuming exponential decay is an appropriate functional form for the dissipation of inner discs at older ages \citep[see discussion in][]{Mamajek09}, for a putative population of 10$^{4}$ Sco-Cen members we may (very) naively expect only 5--10 stars to show signs of accretion at 16--17~Myr  given an $e$-folding time-scale of $\tau=2.3$~Myr \citep{Fedele10}. Even with accretor fractions of 1--2 per cent (consistent with $t\approx\,\,$8--10~Myr) there should exist no more than a few hundred examples across the association. The serendipitous discovery of two such stars in a kinematically-selected survey not looking at Sco-Cen or targeting accretors and infrared excess sources thus seems very surprising, and demands a more nuanced explanation than simple exponential decay at fixed age and decay time. In the following paragraphs we consider two possible solutions to this problem -- significant age spreads within Sco-Cen and stellar mass-dependent disc dissipation. 

 After considering the effects of observational uncertainties, \citet{Mamajek02} concluded that the \emph{intrinsic} 1$\sigma$ age spreads of G-type stars in UCL and LCC are 3~Myr and 2~Myr, respectively. This implies that almost all star formation ceased in the subgroups approximately 5--10~Myr ago. \citet{Pecaut12} found similar median ages for UCL (16~Myr) and LCC (17~Myr) F-type members, but with larger intrinsic age spreads of $\pm$4--7~Myr and $\pm$0--8~Myr, depending on the choice of evolutionary models. In the maximal cases these give 2$\sigma$ limits on the cessation of star formation which are  comparable to the ages of the subgroups themselves. Such age spreads are not unexpected given that star formation in UCL and LCC probably unfolded in a series of small embedded clusters and filaments in a large, dynamically unbound complex \citep{Preibisch08}. Given the large age spreads observed in higher-mass Sco-Cen members, it is therefore plausible that \starshort{205} and \starshort{91} may be as young as 5--10~Myr and among the last generation of stars to form across the region. In addition to ongoing accretion from dusty inner discs, young ages for \starshort{205} and \starshort{91} are also supported by their elevated CMD positions and undepleted photospheric lithium (Fig.~\ref{fig:lithium_scocen}). 
 
A second possibility is that the time-scale for circumstellar disc dispersal is stellar mass dependent, with the processes responsible for clearing optically-thick primordial discs being less efficient around lower-mass stars \citep[e.g.][]{Kennedy09}. Such a trend has been observed in several young groups, including Upper Scorpius \citep{Carpenter06,Luhman12b}. These studies indicate that a significant fraction ($\sim$25 per cent) of inner discs around low-mass stars can survive for at least 10 Myr. There are hints such a phenomenon may extend to the rest of Sco-Cen. Modulo the association mass function, three M-type accretors have now been identified in UCL and LCC, the same number of earlier type members showing signs of accretion. Larger numbers of K and M-type members are necessary to test this hypothesis.  Both \starshort{205} and \starshort{91} have strong near-UV excesses and were included in the spectroscopic sample for this reason. The inclusion of \emph{GALEX} photometry in future Sco-Cen surveys should be a useful way to select additional active and accreting objects \citep[e.g.][]{Rodriguez11,Rodriguez13,Shkolnik11}.

Finally, we comment on the recent suggestion of \citet{Song12} that UCL and LCC are as young as $\sim$10~Myr. This age estimate was based on $\sim$90 GKM stars which showed  levels of lithium depletion bracketed between TWA (8--10~Myr) and the $\beta$~Pictoris Association (Fig.~\ref{fig:lithium_scocen}), which had previously been considered to have an age of $\sim$12 Myr. Given recent upward revisions of the age of $\beta$~Pic to 20--25~Myr \citep{Binks14,Malo14}, the \citet{Song12} results now imply ages between 10 and 20~Myr, consistent with the 16--17~Myr median subgroup ages and age spreads presented by \citet{Pecaut12} and discussed in this section. Again, many more M-type Sco-Cen members  are needed to confirm the lithium depletion age of the association and examine trends with stellar parameters such as rotation and binarity.

\section*{Acknowledgements}

We thank Marc White, Evgenya Shkolnik, David Rodriguez, Mike Ireland, Mark Pecaut, Grant Kennedy and Eric Mamajek for interesting discussions on aspects of this work, and the referee for a thorough and constructive review. SJM gratefully acknowledges the IAU and organisers of IAUS 314 for the award of a travel grant to present parts of this work. This research has made extensive use of the VizieR and SIMBAD databases provided by the Centre de Donn\'ees astronomiques de Strasbourg, the Virtual Observatory SED Analyser \citep[VOSA;][]{Bayo08}, the \textsc{topcat} software tool \citep{Taylor05} and \textsc{astropy}, a community-developed Python package for astronomy \citep{Astropy-Collaboration13}.  

\bibliographystyle{mnras}
\bibliography{paper}

\appendix
\section{Sco-Cen members with \emph{Hipparcos} astrometry and radial velocities}\label{sec:rvcat}

\begin{table*}
\caption{Radial velocities and space motions of Sco-Cen members proposed by \citet{de-Zeeuw99} from the \emph{Hipparcos} catalogue. The final column lists the source of the radial velocity; either \citet{Chen11}, PCRV \citep{Gontcharov06} or CRVAD-2 \citep{Kharchenko07}. The full table with extra columns is available in machine-readable format as Supporting Information with the online version of the article and on the CDS VizieR service.}
\label{table:rvcat}
\begin{tabular}{cccccccccccl}
\hline
  \multicolumn{1}{|c|}{HIP} &
  \multicolumn{1}{c|}{RA} &
  \multicolumn{1}{c|}{Dec.} &
  \multicolumn{1}{c|}{RV} &
  \multicolumn{1}{c|}{$\sigma$(RV)} &
  \multicolumn{1}{c|}{$U$} &
  \multicolumn{1}{c|}{$\sigma(U)$} &
  \multicolumn{1}{c|}{$V$} &
  \multicolumn{1}{c|}{$\sigma(V)$} &
  \multicolumn{1}{c|}{$W$} &
  \multicolumn{1}{c|}{$\sigma(W)$} &
  \multicolumn{1}{c|}{RV source} \\
 & \multicolumn{1}{c|}{(deg)} &
  \multicolumn{1}{c|}{(deg)} &
  \multicolumn{1}{c|}{(\kms)} &
  \multicolumn{1}{c|}{(\kms)} &
  \multicolumn{1}{c|}{(\kms)} &
  \multicolumn{1}{c|}{(\kms)} &
  \multicolumn{1}{c|}{(\kms)} &
  \multicolumn{1}{c|}{(\kms)} &
  \multicolumn{1}{c|}{(\kms)} &
  \multicolumn{1}{c|}{(\kms)} \\
\hline
  50520 & 154.771 & $-$64.676 & 15.8 & 0.7 & $-$8.3 & 0.6 & $-$18.7 & 0.7 & $-$5.1 & 0.3 & PCRV\\
  50847 & 155.742 & $-$66.901 & 12.0 & 4.2 & $-$10.0 & 1.4 & $-$15.8 & 3.9 & $-$3.1 & 0.6 & PCRV\\
  53701 & 164.808 & $-$61.321 & 15.7 & 4.3 & $-$8.8 & 1.8 & $-$19.7 & 4.1 & $-$5.8 & 0.5 & PCRV\\
  55334 & 169.97 & $-$70.618 & 21.3 & 0.9 & $-$6.8 & 0.9 & $-$25.7 & 0.9 & $-$7.8 & 0.3 & Chen\\
  55425 & 170.252 & $-$54.491 & 9.4 & 5.0 & $-$12.7 & 1.8 & $-$15.4 & 4.7 & $-$6.2 & 0.6 & CRVAD2\\
 \dots & \dots & \dots & \dots & \dots & \dots &  \dots &\dots & \dots & \dots & \dots & \dots\\
\hline\end{tabular}
\end{table*}

\begin{figure*} 
   \centering
   \includegraphics[width=0.33\linewidth]{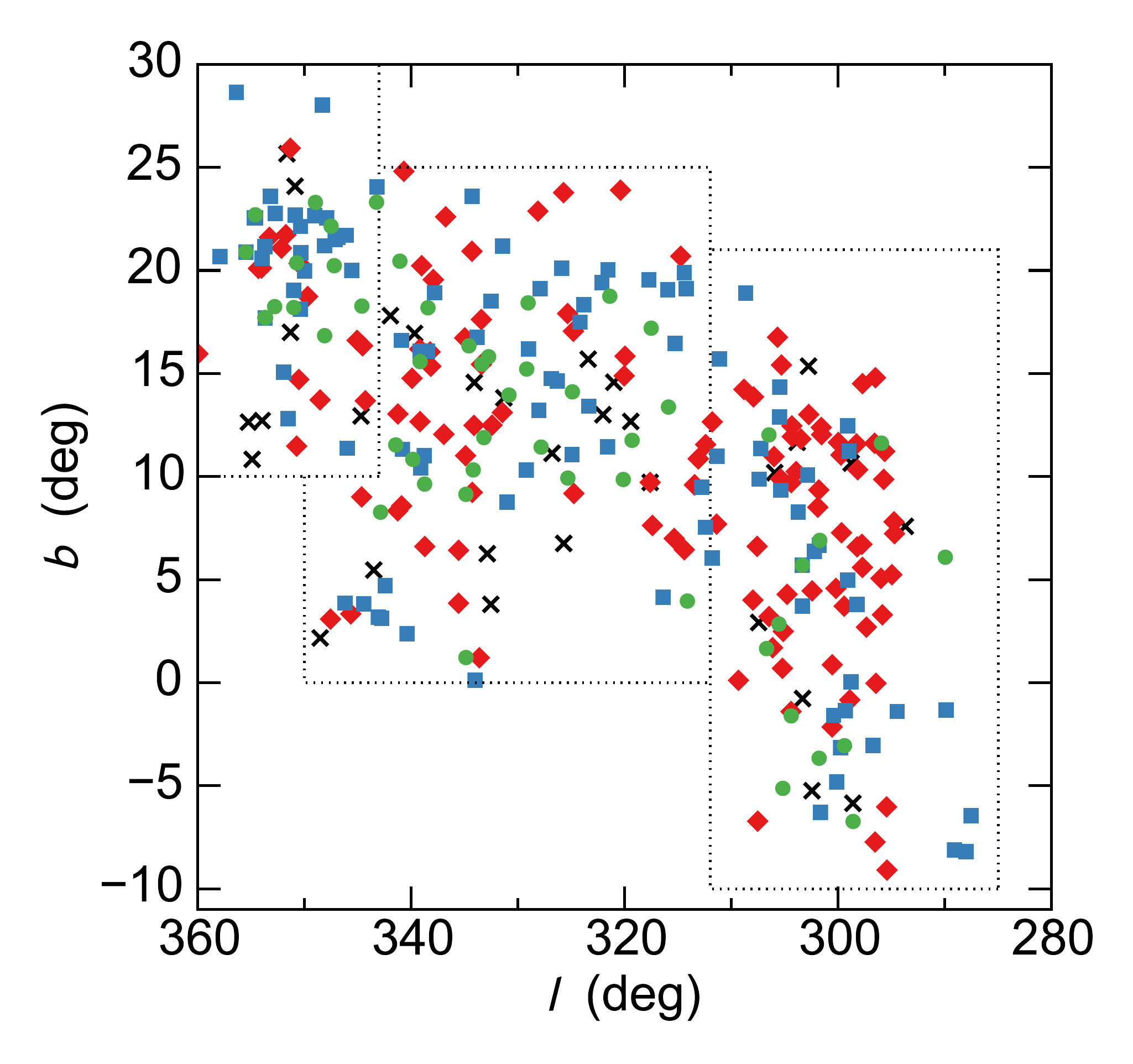} 
   \includegraphics[width=0.33\linewidth]{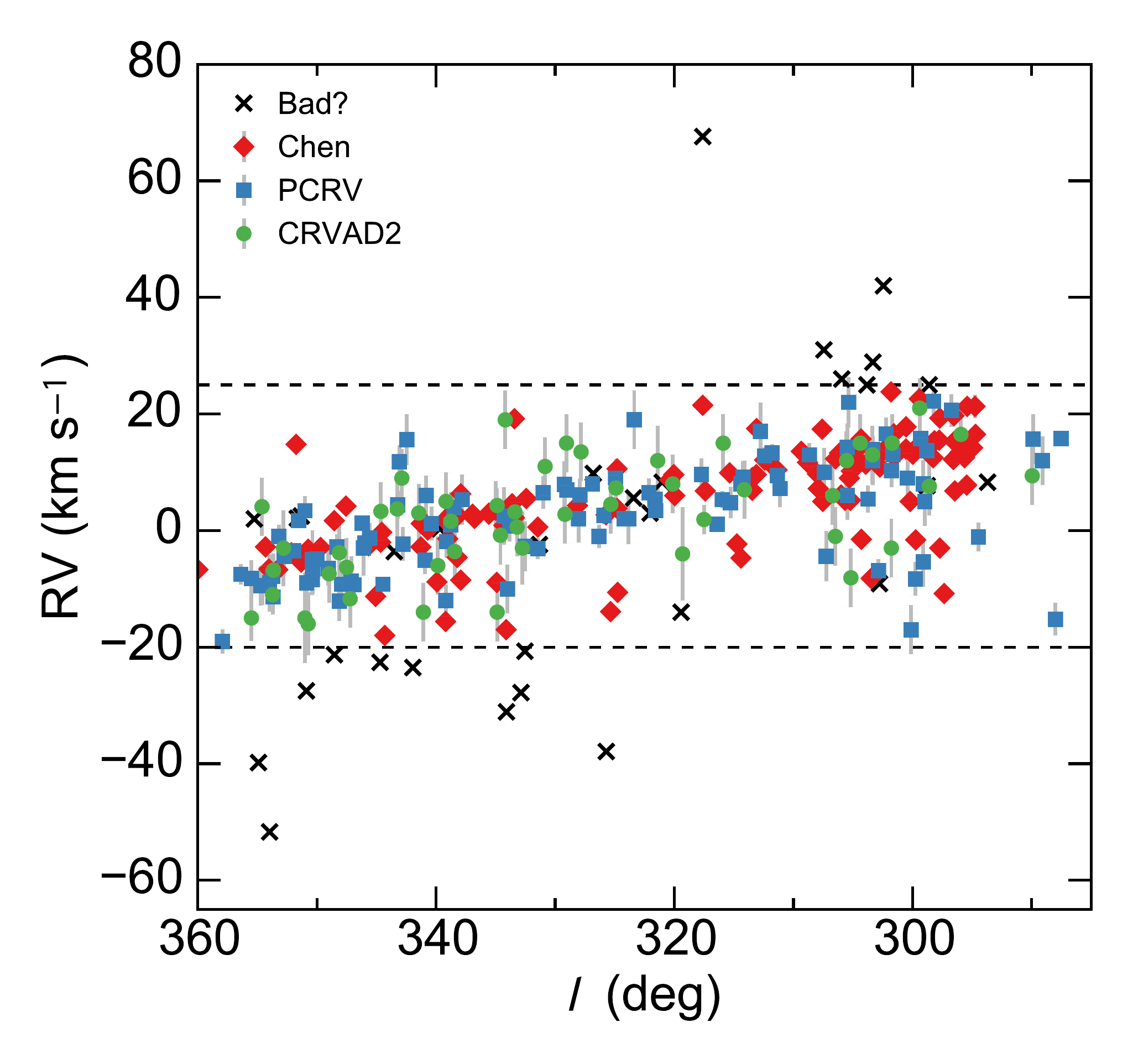} 
   \includegraphics[width=0.33\linewidth]{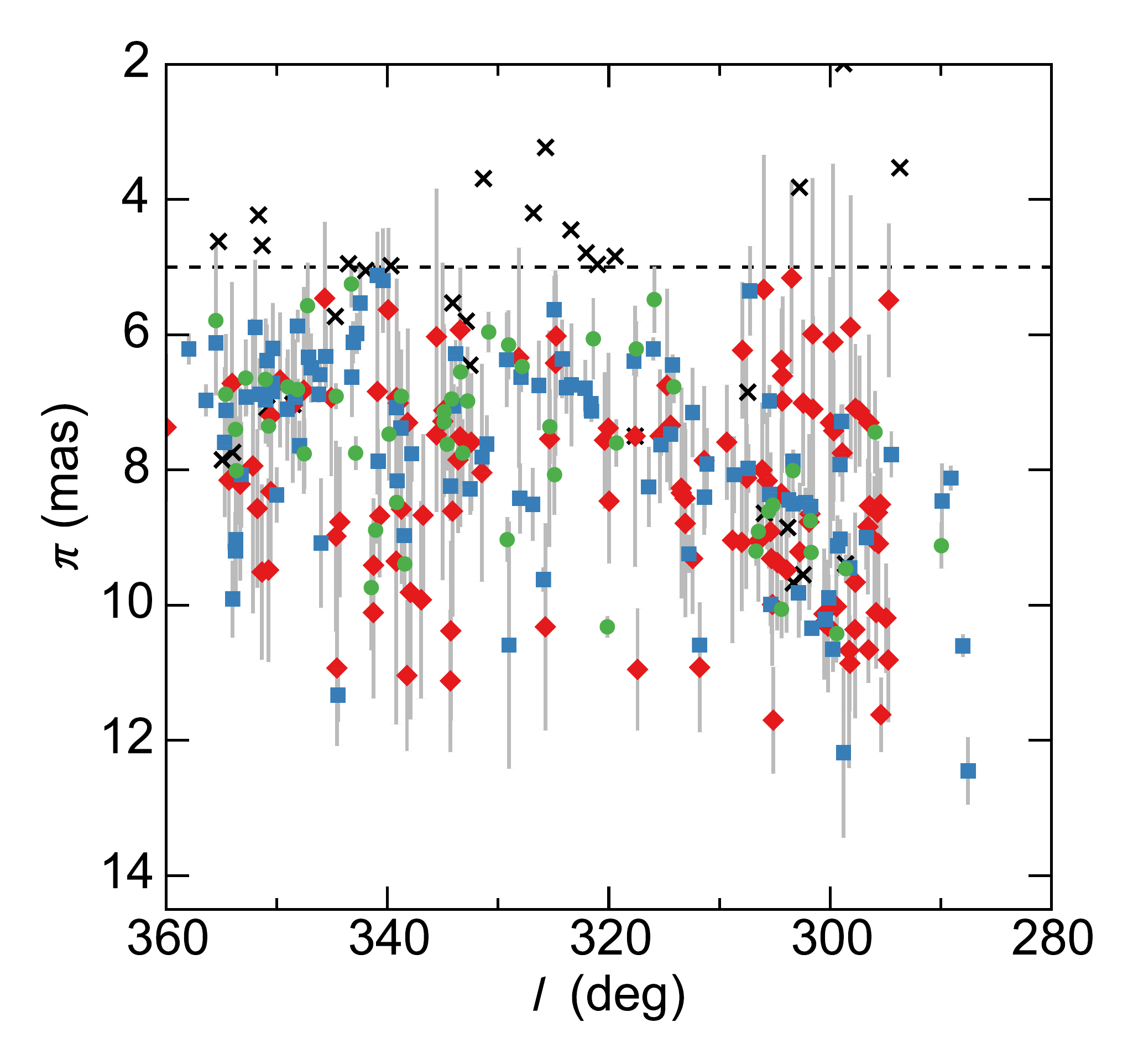} 
      \caption{Position, radial velocity and parallax distributions of the stars in Table\,\ref{table:rvcat}. The source of the RV is indicated: \citet{Chen11} (red diamonds), PCRV (blue squares) or CRVAD2 (green dots). Crosses show the 31 potentially unreliable stars not included in Fig.~\ref{fig:trends}. These have $d>200$~pc or $-20~\kms\ > \text{RV} > 25~\kms$ (dashed lines).}
   \label{fig:rvcat}
\end{figure*}

\citet{Rizzuto11} presented a list of 436 Sco-Cen members from the new \emph{Hipparcos} reduction \citep{van-Leeuwen07}, including 382 stars with radial velocities in the CRVAD-2 compilation \citep{Kharchenko07}, which they used to construct linear models of the Sco-Cen space motion with Galactic longitude. During the present work we discovered that most of the radial velocities in that sample are non-spectroscopic `astrometric' velocities from \citet{Madsen02} which are inappropriate for calculating space motions or memberships. These were apparently sourced from the Pulkovo Compilation of Radial Velocities \citep[PCRV;][]{Gontcharov06}, however they are not in the published PCRV catalogue. They were part of a preliminary list used to construct CRVAD-2 and later removed prior to publication of the PCRV (G.\,A.\,Gontcharov, personal  communication). 

To construct Fig.~\ref{fig:trends} we therefore assembled our own compilation of purely spectroscopic velocities from several large catalogues in the literature. The sample was constructed as follows. We first crossmatched the 521 members proposed by \citet{de-Zeeuw99} (dZ99) against \citet{van-Leeuwen07} and the list of Sco-Cen stars with Magellan/MIKE velocities from \citet{Chen11}. This returned 181 matches. After removing known spectroscopic binaries and non-members 140 stars were retained.  We chose as a secondary source the PCRV, which contains mean absolute velocities for 35,495 \emph{Hipparcos} stars from over 200 publications, including major surveys. This returned 134 matches against dZ99, and we included the 110 stars not already observed by \citeauthor{Chen11} Finally, we crossmatched  dZ99 against CRVAD-2, after removing all objects sourced from the pre-publication PCRV list (\texttt{ICRV=2}). This gave 57 primaries not already in the PCRV or \citeauthor{Chen11} samples. Several stars were found in more than one catalogue, allowing us to eliminate the  spectroscopic binaries HIP~82747 and HIP 73666. The result of these crossmatches was a list of 305 Sco-Cen members with high quality astrometry and radial velocities (median precision 1.5~\kms). The final compilation is listed in Table\,\ref{table:rvcat} and plotted in Fig\,\ref{fig:rvcat}. Note that Fig.~\ref{fig:trends} includes additional cuts to remove 31 stars with unusually large velocities ($-20~\kms\ > \text{RV} > 25~\kms$) or distances ($d>200$~pc) which are less likely to be members. \\

\section*{Supporting Information}

Additional Supporting Information may be found in the online version of this article:\\

\noindent \textbf{Table~\ref{table:rvcat}.} New \emph{Hipparcos} astrometry, literature radial velocities and re-calculated $UVW$ space motions and $XYZ$ positions of 305 Sco-Cen members proposed by \citet{de-Zeeuw99}.\\

\noindent Please note: Oxford University Press are not responsible for the content or functionality of any supporting materials supplied by the authors. Any queries (other than missing material) should be directed to the corresponding author for the article.

\bsp	
\label{lastpage}
\end{document}